\documentclass[twocolumn]{aastex631}

\usepackage{float}
\usepackage{hyperref}
\usepackage{xcolor}
\usepackage{amsmath}
\usepackage{footnote}

\defcitealias{Lustigyaeger2023}{Lustig-Yaeger \& Fu et al. 2023}
\defcitealias{Moran2023_486}{Moran \& Stevenson et al. 2023}
\defcitealias{May2023}{May \& MacDonald et al. 2023}

\begin{document}

\title{JWST COMPASS: The 3-5 Micron Transmission Spectrum of the Super-Earth L 98-59 c}

\author[0000-0003-3623-7280]{Nicholas Scarsdale}
\affiliation{Department of Astronomy and Astrophysics, University of California, Santa Cruz, CA 95064, USA}

\author[0000-0002-0413-3308]{Nicholas Wogan}
\affiliation{NASA Ames Research Center, Moffett Field, CA 94035, USA}

\author[0000-0003-4328-3867]{Hannah R. Wakeford} 
\affiliation{School of Physics, University of Bristol, HH Wills Physics Laboratory, Tyndall Avenue, Bristol BS8 1TL, UK}

\author[0000-0003-0354-0187] {Nicole L. Wallack}
\affiliation{Earth and Planets Laboratory, Carnegie Institution for Science, 5241 Broad Branch Road, NW, Washington, DC 20015, USA}


\author[0000-0003-1240-6844]{Natasha E. Batalha}
\affiliation{NASA Ames Research Center, Moffett Field, CA 94035, USA}

\author[0000-0001-8703-7751]{Lili Alderson} 
\affiliation{School of Physics, University of Bristol, HH Wills Physics Laboratory, Tyndall Avenue, Bristol BS8 1TL, UK}

\author[0000-0002-8949-5956]{Artyom Aguichine}
\affiliation{Department of Astronomy and Astrophysics, University of California, Santa Cruz, CA 95064, USA}


\author[0000-0003-2862-6278]{Angie Wolfgang}
\affiliation{Eureka Scientific Inc., 2452 Delmer Street Suite 100, Oakland, CA 94602-3017}

\author[0009-0008-2801-5040]{Johanna Teske} 
\affiliation{Earth and Planets Laboratory, Carnegie Institution for Science, 5241 Broad Branch Road, NW, Washington, DC 20015, USA}

\author[0000-0002-6721-3284]{Sarah E. Moran}
\affiliation{Department of Planetary Sciences and Lunar and Planetary Laboratory, University of Arizona, Tuscon, AZ, USA}

\author[0000-0003-3204-8183]{Mercedes L\'opez-Morales} 
\affiliation{Center for Astrophysics ${\rm \mid}$ Harvard {\rm \&} Smithsonian, 60 Garden St, Cambridge, MA 02138, USA}

\author[0000-0002-4207-6615]{James Kirk} 
\affiliation{Department of Physics, Imperial College London, Prince Consort Road, London SW7 2AZ, UK}

\author[0000-0001-5253-1987]{Tyler Gordon}
\affiliation{Department of Astronomy and Astrophysics, University of California, Santa Cruz, CA 95064, USA}

\author[0000-0002-8518-9601]{Peter Gao} 
\affiliation{Earth and Planets Laboratory, Carnegie Institution for Science, 5241 Broad Branch Road, NW, Washington, DC 20015, USA}

\author[0000-0002-7030-9519]{Natalie M. Batalha}
\affiliation{Department of Astronomy and Astrophysics, University of California, Santa Cruz, CA 95064, USA}

\author[0000-0003-4157-832X]{Munazza K. Alam}
\affiliation{Space Telescope Science Institute, 3700 San Martin Drive, Baltimore, MD 21218, USA}

\author[0000-0002-4489-3168]{Jea Adams Redai} 
\affiliation{Center for Astrophysics ${\rm \mid}$ Harvard {\rm \&} Smithsonian, 60 Garden St, Cambridge, MA 02138, USA}



\begin{abstract}

We present a JWST NIRSpec transmission spectrum of the super-Earth exoplanet L 98-59 c. This small (R$_p=1.385\pm0.085$R$_\oplus$, M$_p=2.22\pm0.26$R$_\oplus$), warm (T$_\textrm{eq}=553$K) planet resides in a multi-planet system around a nearby, bright (J = 7.933) M3V star. We find that the transmission spectrum of L 98-59 c is featureless at the precision of our data. We achieve precisions of 22ppm in NIRSpec G395H's NRS1 detector and 36ppm in the NRS2 detector at a resolution R$\sim$200 (30 pixel wide bins). At this level of precision, we are able rule out primordial H$_2$-He atmospheres across a range of cloud pressure levels up to at least $\sim$0.1mbar. By comparison to atmospheric forward models, we also rule out atmospheric metallicities below $\sim$300$\times$ solar at 3$\sigma$ (or equivalently, atmospheric mean molecular weights below $\sim$10~g/mol). We also rule out pure methane atmospheres. The remaining scenarios that are compatible with our data include a planet with no atmosphere at all, or higher mean-molecular weight atmospheres, such as CO$_2$- or H$_2$O-rich atmospheres. This study adds to a growing body of evidence suggesting that planets $\lesssim1.5$R$_\oplus$ lack extended atmospheres.


\end{abstract}



\section{Introduction} \label{sec:intro}

The atmospheres of super-Earth ($\sim$1--1.6 R$_\oplus$) exoplanets have been the subject of significant investigation. This class of planet lacks a solar system analog but along with the sub-Neptunes ($\sim$2--3.5 R$_\oplus$) is one of the two most common planet types (for P$<$100~d) in the Galaxy \citep{howard2012, fulton2017, Dattilo_2023}. Over the past decade, several attempts have been made to study their atmospheres using the Hubble Space Telescope (HST), but no molecular features were unambiguously detected \citep{Tsiaras2016_55cnce, dewit2018_trappistwfc3, Wakeford2019, Edwards2021_lhs1140, LibbyRoberts2022_gj1132, barclay_23_arxiv, Zhou2023}. JWST \citep{Gardner2023}, launched in December of 2021, has already begun to revolutionize the study of small planet atmospheres. The Near-InfraRed Spectrograph (NIRSpec) instrument \citep{Jakobsen2022, Birkmann2022}, which we make use of in this study, has been shown to exceed pre-launch expectations for throughput and PSF quality \citep{Boker2023_nirspec}. NIRSpec's G395H mode has already been used to produce high precision transmission spectra of four planets $<$2~R$_\oplus$: GJ 486 b \citepalias{Moran2023_486}, LHS 475 b \citepalias{Lustigyaeger2023}, GJ 1132 b \citepalias{May2023}, and TOI-836 b \citep{alderson2024_arxiv}.

The JWST COMPASS \citep[Compositions Of Mini-Planet Atmospheres for Statistical Study;][PID \#2512]{batalha_teske_compass_proposal} program was devised in pursuit of high precision transmission spectra of small exoplanets. COMPASS is set to obtain NIRSpec G395H transmission spectra of eleven small ($<$3 R$_\oplus$) exoplanets and will make use of a twelfth from another program (GTO program ID \#1224, PI Birkmann) for our full statistical sample. This full sample will include four pairs of planets in the same system, which will enable intra-system comparison. The program aims to survey the demographics of super-Earth and sub-Neptune atmospheres, and to search for a point of transition between primary and secondary atmospheres. It is also designed to catalyze further community efforts to study small exoplanet atmospheres at the population level with JWST. 

COMPASS targets were chosen from a subset of small ($<$3 R$_\oplus$) exoplanets observed by the Transiting Exoplanet Survey Satellite (TESS) and further selected for radial velocity (RV) mass measurements by the Magellan-TESS Survey \citep{Teske2021_mts}. As advocated for by \citet{Batalha2023_selection}, the COMPASS targets were selected by a quantitative ranking function based on planet radius R$_p$, planet insolation F$_p$, and stellar effective temperature T$_\textrm{eff}$. The selection function also weighted targets by T$^{-2}_\textrm{exp}$, with T$_\textrm{exp}$ representing an estimated exposure time to achieve 30ppm precision with NIRSpec G395H (at 4$\mu$m and R=100) in order to avoid selecting targets with host stars too faint for high-precision transmission spectroscopy. 
This standardization and the use of a quantitative sample selection function are optimal for recovering population-level parameters \citep{Batalha2023_selection}, which is COMPASS' ultimate goal.

L 98-59 is a nearby, bright M3V star hosting three transiting planets \citep{Kostov2019, Cloutier2019, Demangeon21}. Two of these (c, R$_p=1.385\pm0.085$R$_\oplus$; and d, R$_p=1.521\pm0.106$R$_\oplus$) will be used as part of the COMPASS statistical sample. Observations of the former are presented here, while observations of the latter are being conducted by GTO program ID \#1224 (PI Birkmann). The system also hosts a smaller inner planet (b; R$_p=0.85\pm0.06$~R$_\oplus$) and a non-transiting candidate with a minimum mass of $\sim$3~M$_\oplus$ \citep{Demangeon21}. The host star and planet c parameters are shown in Table \ref{tab:sysparams}. 

Because of the host star's apparent brightness and the system's planetary multiplicity, L 98-59 has been the subject of several prior theoretical studies. These studies suggest that low mean molecular weight (mmw) atmospheres could be detectable in as few as 1-2 transits \citep{Pidhorodetska_2021}, while heavier atmospheres like those dominated by CO$_2$ and/or SO$_2$ (due to volcanic outgassing) could require 5 or significantly more transits to detect \citep{Pidhorodetska_2021, Seligman2024}. \citet{Fromont24} also identify the planets of this system as likely to be undergoing or have undergone tens or more of Earth oceans' mass of atmospheric escape, which would favor a heavier and correspondingly less feature-rich atmosphere. Observational studies with HST to date have generally favored this scenario (i.e. a heavier atmosphere), having been unable to detect features in transmission for planet b \citep{Zhou2022}, c \citep{barclay_23_arxiv, Zhou2023}, or d \citep{Zhou2023}. For planet c, the subject of this work, HST studies specifically rule out atmospheric metallicities below $\sim$80$\times$ solar.

\begin{table}
    \centering
    \begin{tabular}{|l|c|}
    \hline
    \hline
        Property & Value \\
    \hline
        J mag & 7.933 \\
        R$_\star$ & 0.303$^{+0.026}_{-0.023}$ R$_\odot$ \\
        T$_\textrm{eff}$ & 3415$\pm$135 K \\
        log(g) & 4.86$\pm$0.13 \\
        $[$Fe/H$]$ & -0.46$\pm$0.26 \\
    \hline
        Period & 3.690678 days \\
        R$_p$ & 1.385$^{+0.095}_{-0.075}$ R$_\oplus$ \\
        M$_p$ & 2.22$^{+0.26}_{-0.25}$ M$_\oplus$ \\
        T$_\textrm{eq}$ & 553$^{+27}_{-26}$ K \\
        Density & 4.57$^{+0.77}_{-0.85}$ g/cc \\
        e & 0.103$^{+0.045}_{-0.058}$ \\
        $\omega$ & 261$^{+20}_{-10}$ \\
        Orbital Distance & 0.0304 AU \\
    \hline
    \end{tabular}
    \caption{Parameters of L 98-59 c and its host star; all values from \citet{Demangeon21}.}
    \label{tab:sysparams}
\end{table}

Here, we present the JWST NIRSpec G395H transmission spectrum of L 98-59 c (aka TOI-175.01) from 2.87--5.05$\mu$m. First, we describe the JWST observations of the target (Section \ref{sec:observations}). We then detail our two independent  data reductions (Section \ref{sec:reduction}) and their results (Section \ref{sec:results}). We end with a discussion of the interpretation of our results and a summary in Sections \ref{sec:discussion} and \ref{sec:summary}.

\section{Observations of L 98-59 \MakeLowercase{c}} \label{sec:observations}

We used JWST to acquire two transits of L 98-59 c on 2023 July 9 and 2023 July 13. This short time separation between visits (one orbital period) was deliberately chosen in order minimize the potential impact of stellar variability. Both visit durations were 4.165 hours (effective integration time 3.318 hours), with 1.35 hours in transit and a 2.815 hours out of transit. Both transits were acquired with JWST NIRSpec G395H in Bright Object Time Series (BOTS) mode. We use the full SUB2048 subarray in NRSRAPID readout mode, illuminating both the NRS1 and NRS2 detectors, giving wavelength coverage from 2.87--3.72$\mu$m and 3.82-5.05$\mu$m. Due to the brightness of the host star, both of our observations use four groups per integration to avoid triggering nonlinear behavior in the detector when near saturation, resulting in 3311 total integrations.

\section{Data Reduction} \label{sec:reduction}

To ensure that our results were robust to reduction technique, we independently reduced the data using two pipelines, \texttt{Eureka!} \citep{Bell2022_eureka}\footnote{https://github.com/kevin218/Eureka} and \texttt{ExoTiC-JEDI} \citep{Alderson2022_jedi_zenodo}\footnote{https://github.com/Exo-TiC/ExoTiC-JEDI}. Both are freely-available and open source codes that have been benchmarked against each other and other codes in previous JWST studies (e.g., \citealt{Alderson2023_ers}; \citetalias{Moran2023_486, May2023}). We describe the details of each reduction in the following subsections. 

\subsection{Reduction 1: \texttt{Eureka!}}\label{sec:eurekareduction}

We used the \texttt{Eureka!} (v0.10) package \citep{Bell2022_eureka} for our first reduction. Stages 1 and 2 are wrappers of the same stages of the \texttt{jwst} pipeline \citep[v1.12.2,][]{bushouse_2023_8399938}, with the following modifications. In Stage 1, \texttt{Eureka!} performs line-by-line background subtraction at the group level, which serves to remove ``striping'' caused by 1/$f$ noise \citep{Alderson2023_ers, Rustamkulov2023}.
We use a 10$\sigma$ rejection threshold in the jump step. We also test reductions without the jump step, which can incorrectly flag jumps and correspondingly decrease precision, but find that this increases the number of outlier points in later stages without improving data quality. In Stage 2, we skip the flat field and absolute photometry steps, with the conversion to physical flux units being unnecessary because we only require relative flux to produce a transmission spectrum. 

\texttt{Eureka!} Stage 3 performs optimal spectral extraction \citep{Horne1986} to produce 1-D stellar spectra from the 2-D pixel array for each integration. To do this, the center of the spectral trace in each column is identified and shifted to produce a flattened trace for extraction. We iterate over combinations of aperture half-widths and background exclusion half-widths from 2 to 8 pixels and 6 to 11 pixels, respectively, and select the combination that minimizes the median absolute deviation in the resulting white light curve for each detector. For both visits in this dataset, we select aperture half-widths of 5 and 4 pixels, and background exclusion half-widths of 8 and 7 for NRS1 and NRS2, respectively. In Stage 3, the \texttt{Eureka!} v0.10 default is to mask only the DO\_NOT\_USE (0) flag. We additionally mask (i.e. set to 0) pixels flagged with SATURATED (1), DEAD (10), HOT (11), LOW\_QE (13), and NO\_GAIN\_VALUE (19), which when left unmasked introduce clearly non-astrophysical temporal structure in some spectroscopic light curves. To remove remaining 1/f noise at the integration level, we perform background subtraction again in Stage 3. We used 10 and 60$\sigma$ outlier rejection thresholds for background subtraction and optimal spectral extraction, respectively.
In Stage 4, we divide the spectra into 42 bins across NRS1 and 64 bins across NRS2, corresponding to bin widths of 30 pixels or approximately 0.02 microns. 

After Stage 4, we do not make use of \texttt{Eureka!}'s Stages 5 and 6, which perform light curve fitting and produce the transmission spectrum, respectively. Instead, we use a custom fitting code as described in \citet{wallack2024_arxiv}. In brief, we fit both the white light curves and spectroscopic light curves using the \texttt{emcee} package \citep{ForemanMackey2013}. For each visit and each detector, we first fit each white light curve, fitting for the R$_{p}$/R$_{*}$, inclination, a/R$_{*}$, and T$_{0}$ using the \texttt{batman} package \citep{Kreidberg2015} while fixing the period (3.690678\,days) and eccentricity (0.103) to the values in \citet{Demangeon21}. We treat limb darkening using the quadratic limb darkening coefficients from \texttt{ExoTiC-LD} \citep{Grant2022} and the 3D Stagger model grid \citep{Magic2015}. We use stellar properties of stellar effective temperature T$_\textrm{eff}=$3415K, log surface gravity log(g)=4.86, and metallicity [Fe/H]= -0.46 \citep{Demangeon21}. 
Our instrumental noise model is of the form 

$$ S(\lambda) = s_0 + s_1 \times T + s_2 \times X+ s_3 \times Y \mathrm{,}$$ 

where X and Y are the normalized positions of the trace on the detector, T is time, and $s_i$ are the free parameters in our systematic noise model. 
This systematics model was chosen as the simplest functional form that incorporated time and trace position. We also test models with combinations of sX and sY (the change in detector position) but find these are not preferred by model comparison. We fit the astrophysical parameters and instrumental model to the raw light curves simultaneously along with a per-point error inflation term that is added in quadrature with the errors determined from \texttt{Eureka!}. We do not fit the first 30 minutes of data, which may exhibit stronger or non-linear ramps compared to the rest of the dataset. 
We first fit our joint astrophysical and instrumental noise model with a Levenberg–Marquardt least-squares minimization. We then initialize 3$\times$ the number of free parameters for the MCMC walkers (27 in the white light curve fits) to the results of that Levenberg–Marquardt least-squares minimization. We then utilize 50,000 steps for an initial burn-in followed by 50,000 production steps. When fitting the spectroscopic light curves, we fix the inclination, a/R$_{*}$, and T$_{0}$ from each detector to the median of the MCMC chains from the white light curve fit, leaving the R$_{p}$/R$_{*}$ as the only astrophysical parameter fit. We use the same systematic noise model and aforementioned fitting process in the spectroscopic light curve fits. In all cases, we take the median of each chain as the best-fit value and the standard deviation as the associated error on each parameter.

\begin{figure*}
    \centering
    \includegraphics[width=0.99\textwidth]{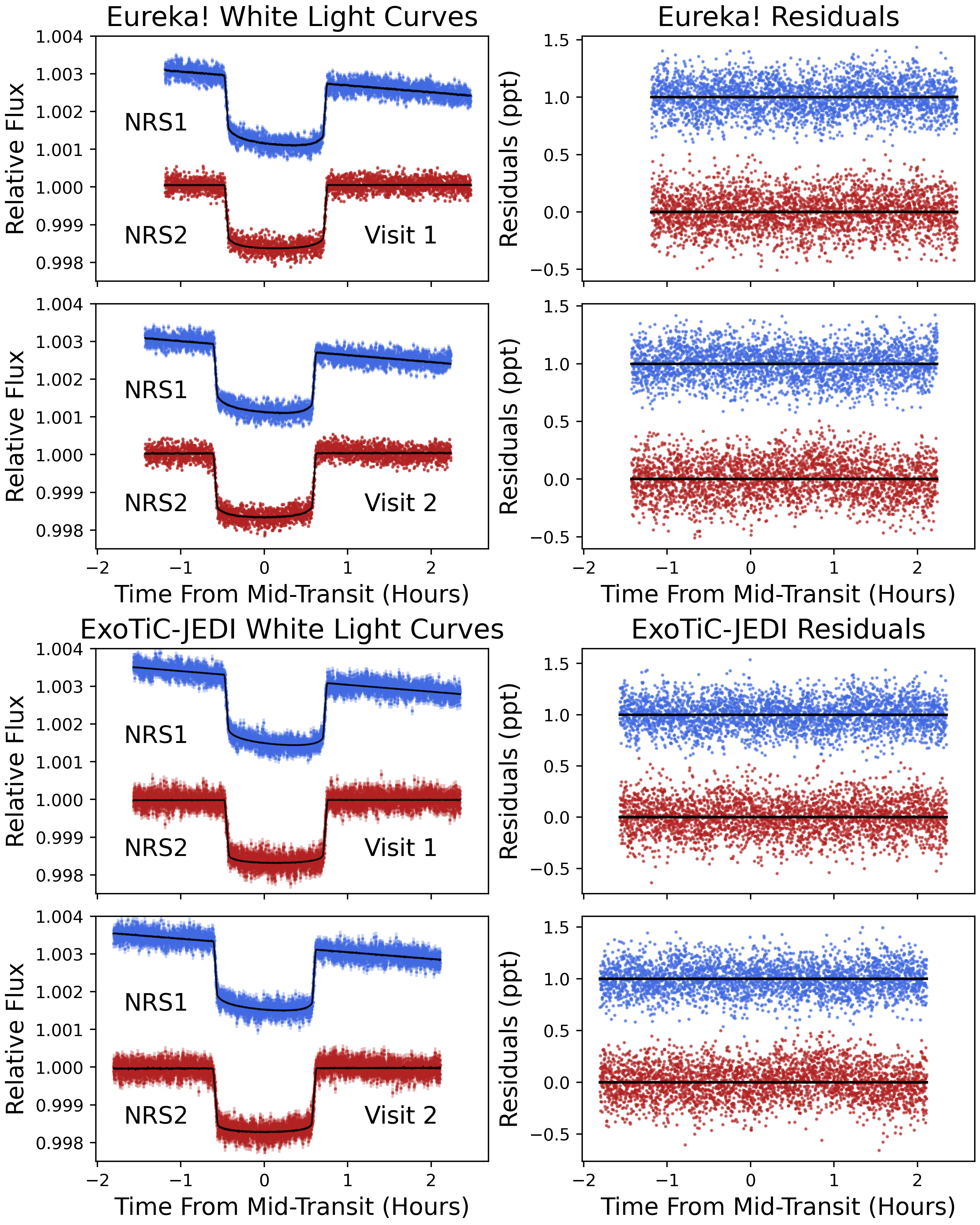}
    \caption{Left column: The L 98-59 c white light curves from each reduction, visit, and detector. Our best-fit models, including systematics, are overplotted on the data in black. Right column: the residuals from the models. 
    }
    \label{fig:combined_wlcs}
\end{figure*}

\begin{figure}
    \centering
    \includegraphics[width=0.49\textwidth]{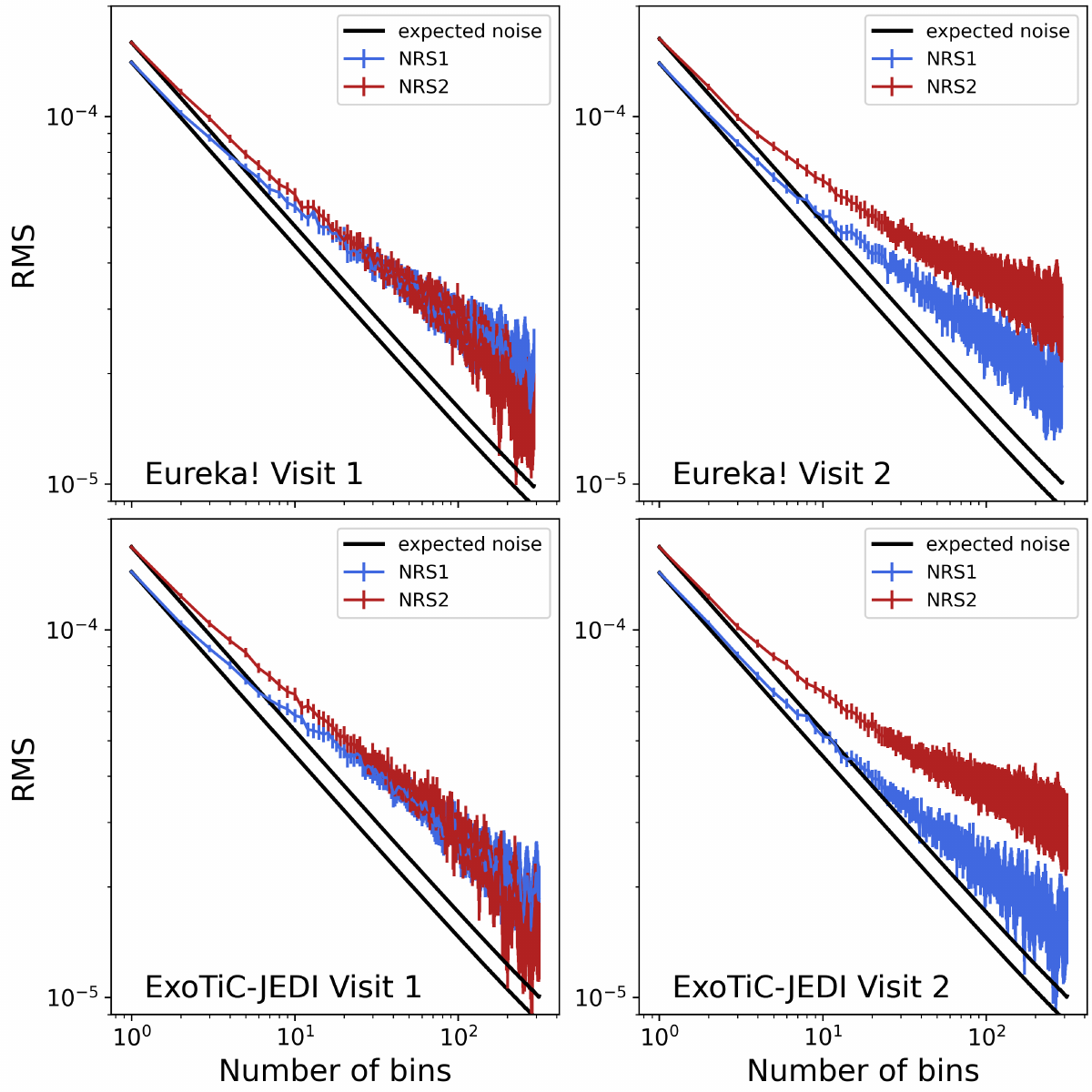}
    \caption{The white light curve Allan deviation plots for each detector and visit from the \texttt{Eureka!} (top) and \texttt{ExoTiC-JEDI} (bottom) reductions. Idealized (i.e. purely Gaussian) noise would fall along the black line in each plot; each detector instead has some amount of red noise, probably due to instrumental systematics that are unaccounted for by our model.}
    \label{fig:allan_combined}
\end{figure}

We also jointly fit the two visits. We treat each detector separately, but fit both NRS1 visits together and both NRS2 visits together. We utilize the same fitting process as described above, but assume common astrophysical parameters between each of the visits for each detector. We fit a common T$_{0}$ where we have normalized the time arrays to the expected time of transit, and common R$_{p}$/R$_{*}$, inclination, and a/R$_{*}$ values. We fit the two detectors separately to determine any offsets due to the differing behaviors of the two detectors. We again utilize a Levenberg–Marquardt least-squares minimization to initialize the chains of an MCMC for the joint fits, initializing 2$\times$ the number of free parameters for the MCMC walkers to the results of a Levenberg–Marquardt least-squares minimization. As with the individual visit fitting, we fix the a/R$_{*}$, inclination, and T$_{0}$ to those from the white light curve of the corresponding detector when performing the spectroscopic light curve fits. Our bins for the spectroscopic light curves from both visits are the same, so we also jointly fit the spectroscopic light curves from both visits. For both the joint white light curves and spectroscopic light curves we again use 50,000 steps for an initial burn-in followed by 50,000 production steps with 2$\times$ the number of free parameters for the number of walkers. The white light curves and Allan deviation plots resulting from this fitting are shown in Figures \ref{fig:combined_wlcs} and \ref{fig:allan_combined}, respectively.

\subsection{Reduction 2: \texttt{ExoTiC-JEDI}}\label{sec:jedireduction}

The Exoplanet Timeseries Characterisation - JWST Extraction and Diagnostic Investigator (\texttt{ExoTiC-JEDI}) package \citep{Alderson2022_jedi_zenodo} performs a full extraction, reduction and analysis of exoplanet transmission spectroscopy time-series data, from JWST \texttt{uncal} files to light curve fitting. We extract and analyse the NIRSpec/G395H NRS1 and NRS2 independently throughout the reduction. \texttt{ExoTiC-JEDI} uses the \texttt{jwst} pipeline tools (v1.8.2) to perform linearity, dark current, and saturation corrections. We set the jump correction threshold to 15 and use a custom destriping routine to remove 1/$f$ noise at the group level using a median of the un-illuminated pixels. For background subtraction, we set a threshold of 15$\sigma$ and use a custom bias routine as described in \citet{Alderson2023_ers}. We use the \texttt{jwst} pipeline to extract the Stage 2 products: 2D wavelength array and exposure times. 

In Stage 3, \texttt{ExoTiC-JEDI} uses the data quality flags from the \texttt{jwst} pipeline. We replace pixels identified as BAD (0), SATURATED (1), DEAD (10), HOT (11), LOW\_QE (13), and NO\_GAIN\_VALUE (19) with the median of the surrounding pixels. After later inspection, we additionally flag pixel [24,1970] in NRS2 as it appeared to be a bad pixel that otherwise escaped correction. Additional bad pixels, often caused by cosmic rays, are identified in the data cube using a 20$\sigma$ threshold in time and 6$\sigma$ threshold spatially. We replace identified pixels with the median of the surrounding 10 integrations and 20 pixels in the row. In total we remove $\approx$\,0.4\% of pixels across the observation due to DQ flags, with a further total of $\approx$\,1,400 pixels (0.0003\%) replaced using our cleaning thresholds. There is often remaining 1/$f$ noise at the integration level, which we remove by masking the illuminated region of the detector and subtracting the median of the unilluminated pixels in each column from each pixel. To extract the 1D stellar spectrum we fit a Gaussian to each column of the data and use a fourth-order polynomial to fit the median filter smoothed trace centers and widths. The trace centers and widths are then used to determine a simple aperture region 5$\times$ the trace full width at half maximum (FWHM) ($\approx$ 6.5 pixels). We use intrapixel extraction to obtain our 1D stellar spectrum. We also measure the trace position movement on the detector in the x- and y-position for detrending at later stages.  

We perform light curve fitting on NRS1 and NRS2 white light curves, as well as spectroscopically across the full wavelength range. Using the white light spectra for each detector and visit, we fit for the planetary system inclination and $a/R_\star$ while fixing the period (3.690678\,days) and eccentricity (0.103) to literature values presented by \citet{Demangeon21}. These parameters, along with the center of transit time, are then held constant in the spectroscopic light curve analysis. We additionally fit for observatory and instrument systematics which we discuss in more detail in the next paragraph. Stellar limb-darkening coefficients are calculated using the \texttt{ExoTiC-LD} package with using a non-linear limb-darkening law and stellar models from Set One of the MPS-ATLAS stellar models \citep{Kostogryz2022, Kostogryz2023}. \texttt{ExoTiC-LD} contains a grid of limb-darkening parameters as a function of 3 variables: stellar temperature, metallicity, and gravity. \texttt{ExoTiC-LD} then uses trilinear interpolation to approximate intermediate values of parameters, where for our case T$\mathrm{eff}$\,=\,3415\,K, logg\,=\,4.86, and  $\mathrm{[Fe/H]}$\,=\,-0.46 \citep{Demangeon21}. Limb darkening values are then fixed throughout our light curve analysis. 


To fit for the transit depth in each bin we use a least-squares optimizer with a \texttt{batman} \citep{Kreidberg2015} transit model. We simultaneously fit a series of systematic models to the data (see \citealt{Alderson2022_jedi_zenodo}) and determine the optimal systematic model based on the negative log-likelihood, which incorporates a penalization in complexity based on the AIC (Akaike Information Criterion). We find that the best systematic model, $S(\lambda)$, corrects for a linear trend in time, $t$, plus the change in x-position, $x_{s}$, multiplied by the absolute magnitude of the y-positional change, $|y_{s}|$, such that $S(\lambda) = s0 + (s1 \times x_{s}|y_{s}|) + (s2 \times t) \mathrm{,}$ where $s0, s1, s2$ are coefficient terms. Our white light curves and corresponding Allan deviation plots are displayed in Figure \ref{fig:combined_wlcs} and Figure \ref{fig:allan_combined}, respectively.

\subsection{A Note on Red Noise and Disagreement Between Best-Fit Astrophysical Parameters}\label{subsec:fitvsfixed}

As can be seen by-eye in Figure \ref{fig:combined_wlcs} and the Allan plots in Figure \ref{fig:allan_combined}, there is significant excess red noise present in our light curves. We attempt a variety of methods to reduce this. As described above, we vary the aperture and background half-widths in our extraction to minimize the median absolute deviation of the extracted data. We perform an alternate set of reductions that skip the jump step in Stage 1, which can misidentify jumps for low group number observations like ours. However, we find that this approach introduces additional outliers to the light curves without improving their red noise properties. Similar red noise levels have been seen previously in other JWST observations of bright stars using low group number (e.g., \citealt{wallack2024_arxiv}, with NIRSpec; \citealt{ Kirk2024_preprint}, with NIRCam). 

\begin{figure}[h]
    \centering
    \includegraphics[width=0.49\textwidth]{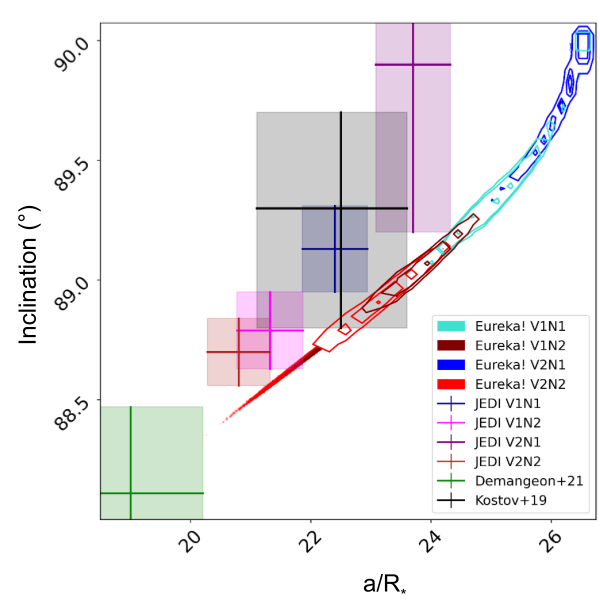}
    \caption{The posteriors from the white light curve fitting to the \texttt{Eureka!} reductions of the transits of L 98-59 c, alongside the best fit values from \texttt{ExoTiC-JEDI}. Note that the behavior of the inclination and a/R$_*$ are discrepant between NRS1 and NRS2, as well as between visits, which is probably unphysical.}
    \label{fig:astrophysparams_corner}
\end{figure}

Probably as a result of this excess red noise, as is visible in Figure \ref{fig:astrophysparams_corner}, several astrophysical parameters resulting from our fits to the white light curves are inconsistent between detectors and visits. For example, in Visit 2, the \texttt{Eureka!} white light curves for NRS1 and NRS2 have best-fit a/R$_\star$ values that are inconsistent with each other at $\sim$2$\sigma$ (Table 2). The \texttt{ExoTiC-JEDI} white light curves for the same visit display a similar discrepancy in a/R$_\star$ (Table 2).
The inclination values in visit 2 are also somewhat discrepant between NRS1 and NRS2.
The first visit a/R$_\star$ and inclination values for the two detectors are consistent with each other at roughly the 1$\sigma$ level for each reduction, but not consistent between reductions (Figure \ref{fig:astrophysparams_corner}). 

This disagreement between parameters also seen in the literature \citep[e.g.,][who report a/R$_\star$ values 22.5$^{+1.1}_{-1.4}$ and 19.0$^{+1.2}_{-0.8}$, in tension at roughly 1.5$
\sigma$]{Kostov2019, Demangeon21}, but it is unexpected to see this tension reproduced between two detectors observing the same transit. This is probably attributable to red noise present in the light curves, which is most pronounced in visit 2 (Figure \ref{fig:allan_combined}), where the disagreement is also most pronounced. In order to ensure that this disagreement does not affect our conclusions, we perform additional white and spectroscopic light curve fitting with the inclination values fixed to 89.3$^\circ$ \citep{Kostov2019}, which is a better match in parameter space to the values obtained from our fitting than the inclination from \citet{Demangeon21}. In this case, the resulting a/R$_\star$ values are consistent between all visits, detectors, and reductions (Table 3). We also attempt fits with both the inclination and a/R$_\star$ fixed to literature values from \citet{Demangeon21}, but the resulting white light curve fits are clearly poor (see Appendix \ref{sec:appendix-astro-param}), with apparent residual structure near ingress and egress (though even in this case, the resulting spectrum is consistent with other techniques). Because it presents an astrophysically consistent picture of the system, we select the fixed inclination model as the fiducial one for generating our transmission spectrum, although we reiterate that fixing and fitting for inclination yield nearly identical spectra. 

\begin{table*}[htbp]
    \centering
    \begin{tabular}{|c|c|c|c|c|c|c|}
    \hline
    \makesavenoteenv{table*}
     & & & R$_{p}$/R$_{*}$ & T$_0$ (BJD)$^\dag$ & a/R$_*$ & i ($^\circ$) \\
    \hline
    \texttt{Eureka!} & Visit 1 & NRS1 & $0.04025\pm1.0\times10^{-4}$ & $-0.003438\pm2.4\times10^{-5}$ & $25.12\pm0.73$ & $89.35\pm0.22$ \\
     &   & NRS2 & $0.03995\pm1.0\times10^{-4}$ & $-0.003329\pm2.8\times10^{-5}$ & $23.82\pm0.64$ & $89.05\pm0.14$ \\
     & Visit 2 & NRS1 & $0.03976\pm8.6\times10^{-5}$ & $-0.003414\pm2.3\times10^{-5}$ & $26.08\pm0.64$ & $89.68\pm0.20$ \\
     &   & NRS2 & $0.04028\pm1.1\times10^{-4}$ & $-0.003472\pm3.0\times10^{-5}$ & $23.07\pm0.68$ & $88.90\pm0.14$ \\
     & Joint Fit & NRS1 & $0.04008\pm5.7\times10^{-5}$ & $-0.003425\pm2.3\times10^{-5}$ & $24.81\pm0.20$ & $89.276\pm0.048$ \\
     &   & NRS2 & $0.03990\pm7.4\times10^{-5}$ & $-0.003392\pm2.0\times10^{-5}$ & $24.47\pm0.37$ & $89.207\pm0.085$ \\
     \hline
     \texttt{ExoTiC-JEDI} & Visit 1 & NRS1 & $0.04046\pm1.0\times10^{-4}$ & $0.005748\pm2.3\times10^{-5}$ & $22.42\pm0.54$ & $89.13\pm0.18$ \\
     &   & NRS2 & $0.04013\pm1.0\times10^{-4}$ & $0.005873\pm2.8\times10^{-5}$ & $21.32\pm0.55$ & $88.79\pm0.16$ \\
     & Visit 2 & NRS1 & $0.03990\pm1.0\times10^{-4}$ & $0.000257\pm2.3\times10^{-5}$ & $23.71\pm0.62$ & $89.98\pm0.70$ \\
     &   & NRS2 & $0.04040\pm9.8\times10^{-5}$ & $0.000199\pm7.9\times10^{-6}$ & $20.83\pm0.52$ & $88.66\pm0.14$ \\
     \hline
     \multicolumn{7}{l}{\vtop{\hbox{\strut $^\dag$Time from respective visit 1 and visit 2 expected mid-transit times BJD 60134.60851 and 60138.29919 (\texttt{Eureka!}) and}\hbox{\strut 60134.59931 and 60138.29550 (\texttt{ExoTiC-JEDI}).}}}
    \end{tabular}
    \label{tab:wlc_free}
    \caption{The best-fit values from the white-light curve fits for L 98-59 c, fitting for all the values.}
\end{table*}

\begin{table*}[htbp]
    \centering
    \begin{tabular}{|c|c|c|c|c|c|c|}
    \hline
    \makesavenoteenv{table*}
     & & & R$_{p}$/R$_{*}$ & T$_0$ (BJD)$^\ddag$ & a/R$_*$ & i ($^\circ$) \\
    \hline
    \texttt{Eureka!} & Visit 1 & NRS1 & $0.04014\pm1.0\times10^{-4}$ & $-0.003435\pm2.1\times10^{-5}$ & $24.883\pm0.024$ & $89.3$ \\
     &   & NRS2 & $0.03978\pm8.4\times10^{-5}$ & $-0.003330\pm2.7\times10^{-5}$ & $24.882\pm0.025$ & $89.3$ \\
     & Visit 2 & NRS1 & $0.03984\pm7.2\times10^{-5}$ & $-0.003418\pm2.3\times10^{-5}$ & $24.882\pm0.021$ & $89.3$ \\
     &   & NRS2 & $0.04006\pm8.3\times10^{-5}$ & $-0.003442\pm2.8\times10^{-5}$ & $24.871\pm0.026$ & $89.3$ \\
     \hline
     \texttt{ExoTiC-JEDI} & Visit 1 & NRS1 & $0.04046\pm7.0\times10^{-5}$ & $0.005757\pm2.3\times10^{-5}$ & $24.915\pm0.02$ & $89.3$ \\
     &   & NRS2 & $0.04001\pm8.3\times10^{-5}$ & $0.005883\pm2.7\times10^{-5}$ & $24.916\pm0.024$ & $89.3$ \\
     & Visit 2 & NRS1 & $0.04007\pm7.0\times10^{-5}$ & $0.000257\pm2.3\times10^{-5}$ & $24.910\pm0.021$ & $89.3$ \\
     &   & NRS2 & $0.04024\pm8.2\times10^{-5}$ & $0.000240\pm2.6\times10^{-5}$ & $24.892\pm0.024$ & $89.3$ \\
     \hline
     \multicolumn{7}{l}{\vtop{\hbox{\strut $^\ddag$Time from respective visit 1 and visit 2 expected mid-transit times BJD 60134.60851 and 60138.29919 (\texttt{Eureka!}) and}\hbox{\strut 60134.59931 and 60138.29550 (\texttt{ExoTiC-JEDI}).}}}
    \end{tabular}
    \label{tab:wlc_fixi}
    \caption{As Table 2, but with the inclination fixed to 89.3$^\circ$ \citep{Kostov2019}.}
\end{table*}

\section{Transmission Spectrum and Results}\label{sec:results}

\begin{figure*}
    \centering
    \includegraphics[width=0.99\textwidth]{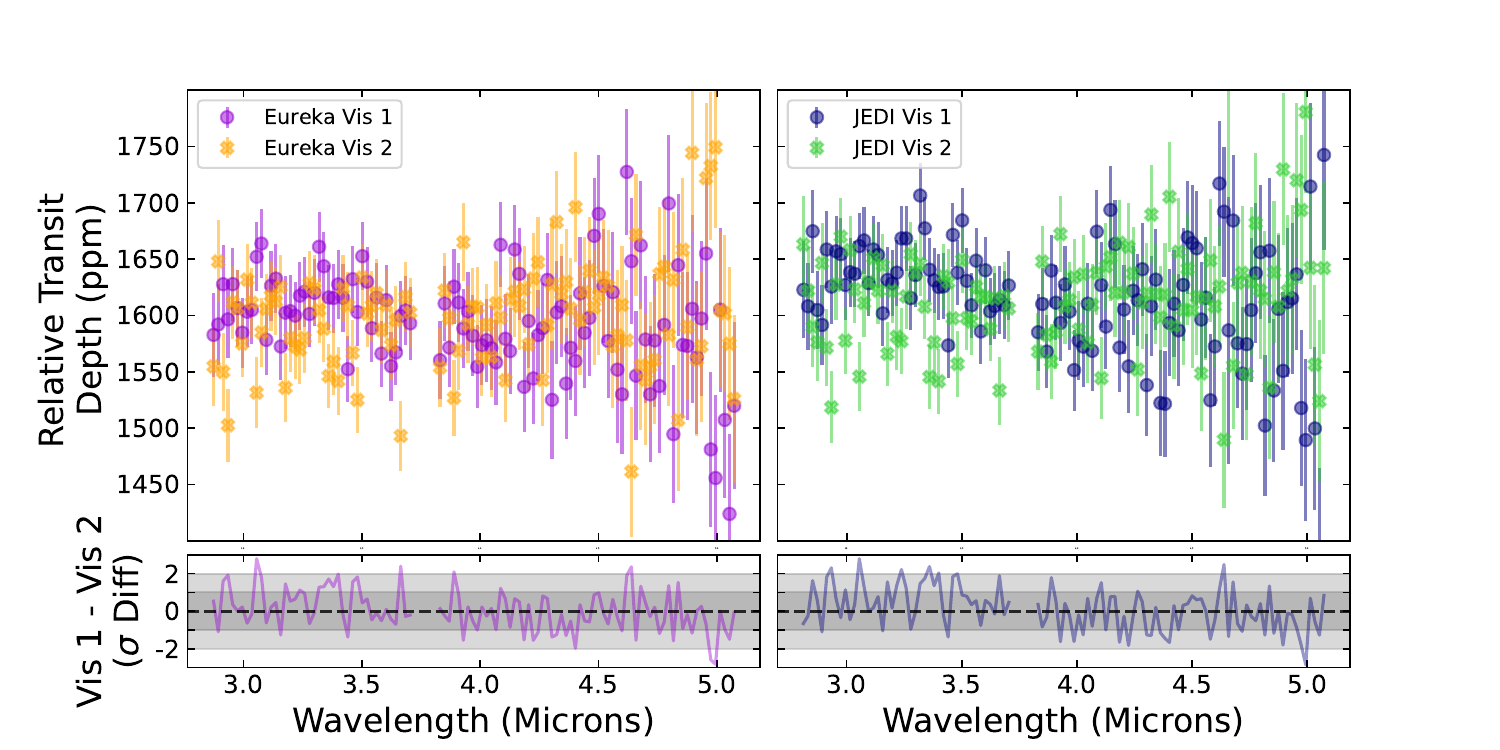}
    \caption{Top: the raw (no offsets applied) transmission spectrum for the \texttt{Eureka!} (left) and \texttt{ExoTiC-JEDI} (right) reductions of L 98-59 c. For each reduction, we show the two visits together. Bottom: The difference between the reductions expressed relative to the pointwise error. The visits are broadly consistent with each other for both reductions, though a handful of points fall outside of 2$\sigma$ consistency. The median difference between visits is 4ppm and 13ppm for \texttt{Eureka!} and \texttt{ExoTiC-JEDI}, respectively.}
    \label{fig:spectra_by_visit}
\end{figure*}

\begin{figure*}
    \centering
    \includegraphics[width=0.99\textwidth]{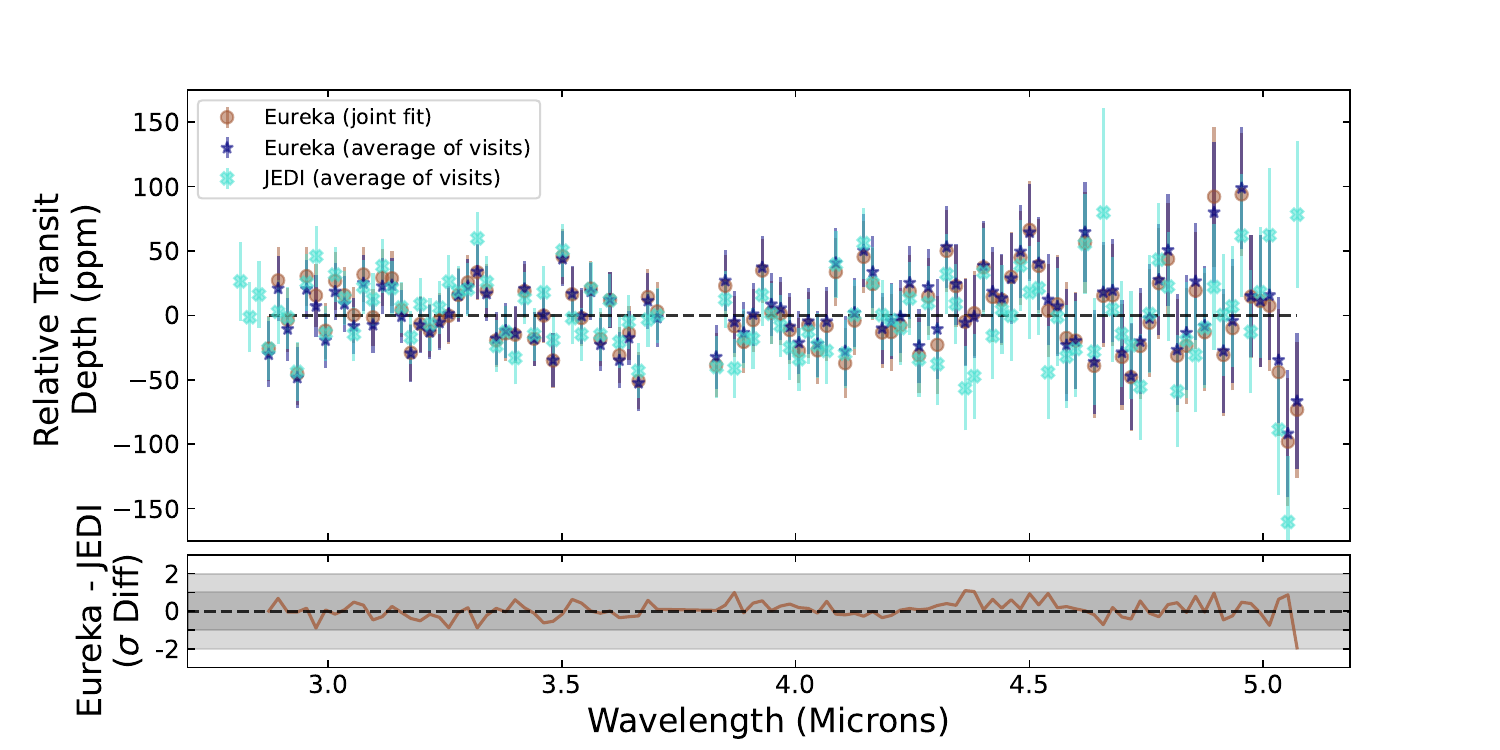}
    \caption{Top: similar to Figure \ref{fig:spectra_by_visit}, but we plot instead the combined spectra from each reduction. For \texttt{Eureka!}, we have combined the spectra by performing a joint transit fit to the two visits simultaneously; for \texttt{ExoTiC-JEDI}, we have created an error-weighted average of the spectra from each of the two visits. Bottom: the difference between the reductions. In this case, several non-edge points become inconsistent at the 1$\sigma$ level, but as we show in Table \ref{tab:nonphys_results}, these small differences do not significantly affect our interpretation of the results.}
    \label{fig:spectra_combined}
\end{figure*}

\begin{figure}
    \centering
    \includegraphics[width=0.49\textwidth]{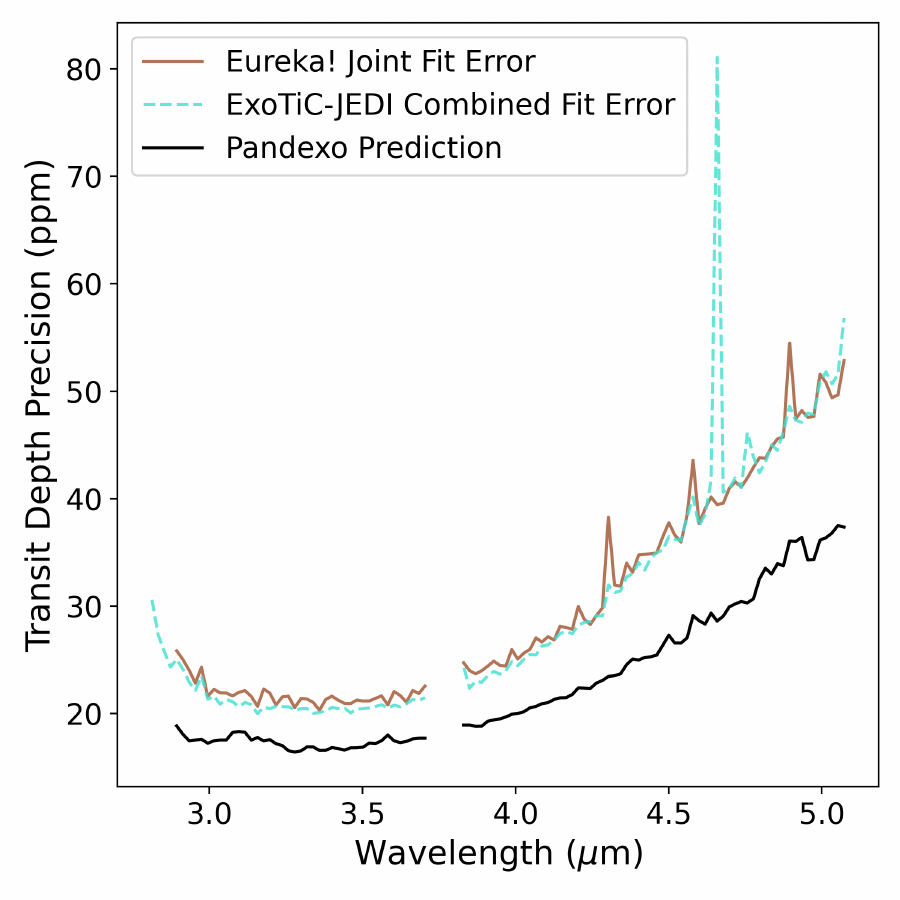}
    \caption{Observed transit depth error by wavelength bin for our two reductions are shown compared to \texttt{pandexo} predictions. \texttt{Eureka!} exceeds the \texttt{pandexo} prediction by a median factor of 1.3 (range 1.1-1.6), and \texttt{ExoTiC-JEDI} by a median factor of 1.3 (range 1.1-2.8).}
    \label{fig:pandexo}
\end{figure}

The 2.8--5.1$\mu$m transmission spectrum of L 98-59 c is shown in Figure \ref{fig:spectra_by_visit} (individual visits) and Figure \ref{fig:spectra_combined} (combined spectra). The \texttt{Eureka!} visits are combined via a joint fit, while the \texttt{ExoTiC-JEDI} vists are combined via an error-weighted average. In these combined spectra, we achieve respective median errors in NRS1 and NRS2 of 22 ppm and 36 ppm for both \texttt{Eureka!} and \texttt{ExoTiC-JEDI}. These are a median factor of 1.3$\times$ higher than post-launch predictions from \texttt{pandexo} \citep{Batalha_2017, Batalha2022}. The two reductions perform comparably, with slightly better performance by \texttt{ExoTiC-JEDI} in NRS1 (Figure \ref{fig:pandexo}). On visual inspection, while the transmission spectra from the individual visits show potential wavelength-dependent variation, they are neither consistent with known molecular features nor consistent between visits. 

\begin{table*}[htbp]
    \centering
    \begin{tabular}{|c|c|c|c|}
        \hline
        \makesavenoteenv{table*}
        Dataset & Preferred Nonphysical Model & Bayes Factor vs. Flat Line & Model Value(s)\footnote{The ppm value of the best fit flat line or step function lines} \\

        \hline

        \texttt{Eureka!} visit 1 & Step & 2.0 (very weak) & 1613, 1593 \\

         \texttt{Eureka!} visit 2 & Flat Line & - & 1598 \\

         \texttt{Eureka!} combined (avg w/ offset) & Flat Line & - & 1591 \\

         \texttt{Eureka!} combined (joint fit) & Flat Line & - & 1602 \\

         \texttt{ExoTiC-JEDI} visit 1 & Step & 481 (strong) & 1642, 1610\\

         \texttt{ExoTiC-JEDI} visit 2 & Flat Line & - & 1612 \\

         \texttt{ExoTiC-JEDI} combined (avg) & Flat Line & - & 1617 \\

         \texttt{ExoTiC-JEDI} combined (avg w/ offsets) & Flat Line & - & 1619 \\
         
         \hline
        
%
%
         %
%
%
%
%

         \hline
    \end{tabular}
    \caption{The full list of reductions and corresponding models preferred by the non-physical testing we describe. We show these results for spectra produced using both fit and fixed astrophysical parameters (see Section \ref{subsec:fitvsfixed}). "Flat Line" means the preferred model is a single constant transit depth, b. "Step" means the preferred model has a distinct depth for the two detectors NRS1 (b1) and NRS2 (b2).}
    \label{tab:nonphys_results}
\end{table*}

\subsection{Feature Detection} \label{sec:feature_detection}

To understand the atmospheric implications of the JWST spectrum, we first search for features by fitting non-physical models to the data. Calculations nominally consider the \texttt{Eureka!} reduction, but we perform the same analyses on the \texttt{ExoTiC-JEDI} dataset as a sensitivity test. We note that performing these tests for both the visits individually and in combination is needed because, as recently emphasized in \citet{alderson2024_arxiv}, even if individual data points agree within 1$\sigma$ (which is approximately true for the combined visits of L 98-59 c), they can yield different best-fit models. Thus, for reproducibility and repeatability it is important to consider at least two reductions in the theoretical interpretation. 

\begin{figure*}
    \centering
    \includegraphics[width=\textwidth]{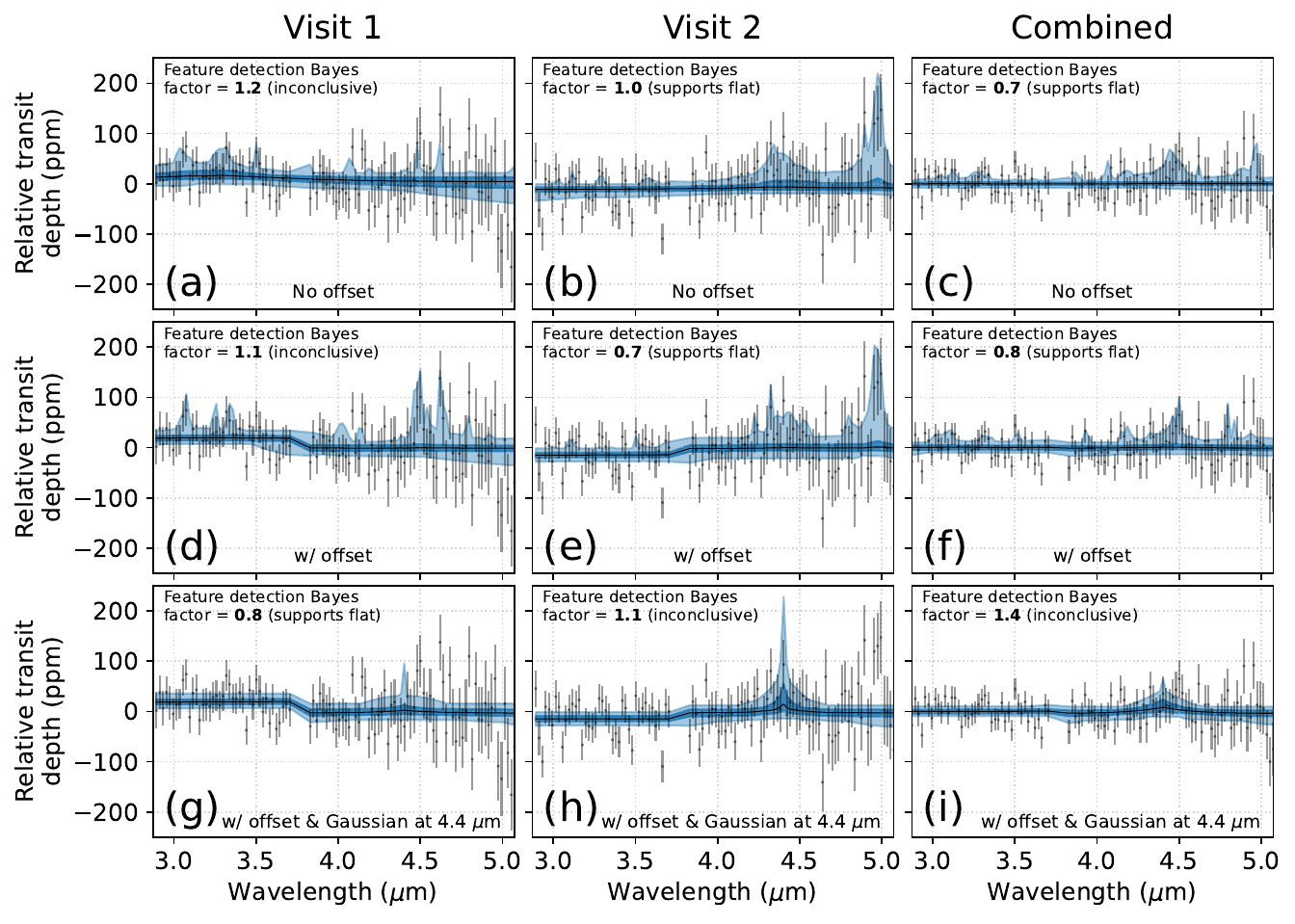}
    \caption{The feature detection significance for visit 1, visit 2 and the combined spectrum from our \texttt{Eureka!} reduction. All panels show the JWST data (black dots) along with the 1-$\sigma$ (dark blue) and 3-$\sigma$ (light blue) posterior distributions from simple Bayesian modeling. Panels (a), (b) and (c) show fits with a Gaussian model representing an arbitrary molecular feature. Panels (d), (e) and (f) consider a Gaussian model but also allow for a vertical offset between the NRS1 ($\lambda < 3.78$ $\mu$m) and NRS2 ($\lambda > 3.78$ $\mu$m) detectors. Panels (g), (h) and (i) fix the mean of the Gaussian curve the 4.4 $\mu$m CO$_2$ feature, again allowing for an offset between NRS1 and NRS2. Each panel reports a ``feature detection Bayes factor'', which we derive by comparing the Gaussian results to featureless Bayesian models (see the main text for details). Overall, this analysis does not strongly detect a feature in any data set.}
    \label{fig:feature_detection}
\end{figure*}

Following previous works \citepalias[e.g.,][]{May2023,Moran2023_486}, we approximate an agnostic spectral feature with a Gaussian curve of the form:

\begin{equation} \label{eq:gauss}
    f(\lambda,A,\mu,\sigma_m,b) = A \exp \left(-\frac{(\lambda - \mu)^2}{2 \sigma_m^2}\right) + b
\end{equation}
Here, $\mu$, $\sigma_m$ and $A$ are the center, standard deviation, and amplitude of a Gaussian, respectively, while $b$ is a vertical offset applied to the entire spectrum and $\lambda$ is wavelength in $\mu$m. We fit this simple model to the JWST spectra using a nested sampling algorithm \citep{MLFriends2019} assuming wide uniform priors for $\mu$ and $b$, and log-uniform priors for $A$ and $\sigma_m$. Figure \ref{fig:feature_detection}a - \ref{fig:feature_detection}c shows the resulting 1-$\sigma$ (dark blue) and 3-$\sigma$ (light blue) posterior distribution for the visit 1, 2 and combined \texttt{Eureka!} spectra. The visit 1 posterior (Figure \ref{fig:feature_detection}a) is inconclusive as to the presence of Gaussian features between $4$ and $5.1$ $\mu$m, while the visit 2 and the joint fit calculations (Figure \ref{fig:feature_detection}b and \ref{fig:feature_detection}c) are comparatively flat. 

To determine the robustness of possible features, we first perform Bayesian model comparison between the Gaussian model (Equation \eqref{eq:gauss}) and a one parameter model that fits the data with a flat line. Each panel in Figure \ref{fig:feature_detection} reports a Bayes factor, which is the Bayesian evidence of the Gaussian model divided by the evidence of a flat-line model. Feature detection Bayes factors less than 1 support a flat-line while values of $B < 3$, $B \sim 3$, $B \sim 12$, and $B \gtrapprox 150$ can be interpreted as inconclusive, weak, moderate, and strong preference for a Gaussian model, respectively \citep{Trotta2008}. The visit 2 and the combined spectra (Figure \ref{fig:feature_detection}b and \ref{fig:feature_detection}c) have feature detection Bayes factors of $B < 1$ indicating preference for a flat line model. On the other hand, the visit 1 Bayes factor ($B = 1.2$; Figure \ref{fig:feature_detection}a) shows only ``inconclusive'' support for a spectral feature.




Figures \ref{fig:feature_detection}d, \ref{fig:feature_detection}e and \ref{fig:feature_detection}f show the results of further Gaussian fits to the data that have an additional parameter allowing the NRS1 data ($\lambda < 3.78$ $\mu$m) to shift relative to NRS2 ($\lambda > 3.78$ $\mu$m). For these experiments the Bayes factors are either inconclusive or showpreference for a featureless model (with a detector offset) over a Gaussian model (with a detector offset) for visit 1, visit 2 and the joint-fit \texttt{Eureka!} spectrum.

We have attempted additional tests which, as a whole, do not support the presence of a feature. Repeating the Figure \ref{fig:feature_detection}d - \ref{fig:feature_detection}f calculations with the \texttt{ExoTiC-JEDI} reduction also indicates preference for a flat line with a detector offset over a model that includes a Gaussian curve. The preferred models for each dataset are summarized in Table \ref{tab:nonphys_results}. Furthermore, we have tried tests that fix the center of the Gaussian curve at the 4.4 $\mu$m CO$_2$ feature, again permitting a detector offset (Figure \ref{fig:feature_detection}g - \ref{fig:feature_detection}i). In this case, no visit or reduction shows a notable preference for a Gaussian feature. Additionally, we found that no feature was favored for a similar fixed-Gaussian test that instead centered the feature at 3.3 $\mu$m where CH$_4$ absorbs.

Overall, this nonphysical model analysis suggests the JWST spectrum is featureless. Regardless, the data still contain substantial information because they can rule out a large swath of atmospheric compositions, which we address in the following subsection.



\subsection{Comparison with Atmospheric Models}

To determine the atmospheres ruled out by the featureless spectrum, we compare the data to a wide range of simulated compositions. We consider gas-combinations at chemical equilibrium at various atmospheric metallicities, as well as simple H$_2$-CO$_2$ and H$_2$-H$_2$O mixtures, all being rough proxies for atmospheric mean molecular weight. Calculations also consider a range of ``opaque pressure levels'', which can be thought of as the top of an optically thick cloud or, alternatively, the total surface pressure. Following past work \citep[e.g.,][]{alderson2024_arxiv, wallack2024_arxiv}, we adopt a simple 1D parameterized pressure-temperature (P-T) profile for all simulations \citep{Guillot2010}. A more sophisticated, self-consistent P-T profile is not warranted because of the lack of clear spectral features. Given the P-T profile, a chemical composition, and an ``opaque pressure level'', we simulate the transmission spectrum with the \texttt{PICASO} code \citep{Batalha2019_picaso} and fit it to the data permitting an offset between the NRS1 and NRS2 detectors. Finally, we perform a $\chi^2$ test to determine the degree to which the model is ruled out by the data.

\begin{figure*}
    \centering
    \includegraphics[width=\textwidth]{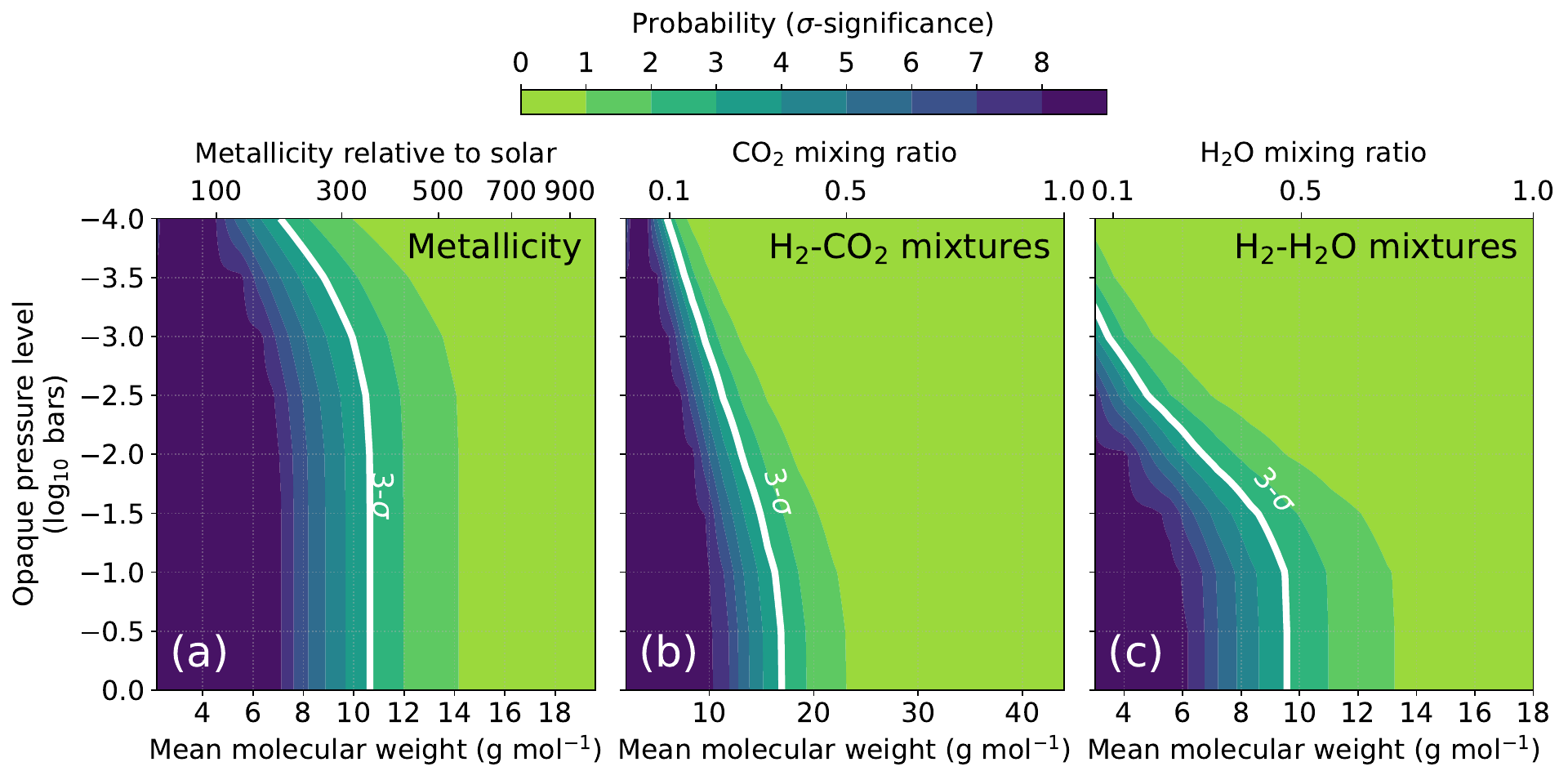}
    \caption{The compatibility of the combined \texttt{Eureka!} spectrum with a wide range of atmospheric compositions and opaque pressure levels. Colored contours show the $\sigma$-confidence to which each model can be ruled out by the data. Model spectra permit an offset between the NRS1 and NRS2 detectors as to best fit the data. Panel (a) considers a grid of atmospheres with compositions between $1 \times$ and $1000 \times$ solar. Panel (b) explores H$_2$-CO$_2$ atmospheres, while panel (c) concerns H$_2$-H$_2$O atmospheres. For an opaque pressure level of 1 bar, the data confidently ($>3$-$\sigma$) rules out atmospheres with mean molecular weights $\lesssim 10$ g mol$^{-1}$ in all cases. Smaller mean molecular weights are allowed by the data for lower opaque pressure levels (e.g., $10^{-4}$ bar).}
    \label{fig:sweep}
\end{figure*}

Figure \ref{fig:sweep}a shows the metallicities and opaque pressure levels incompatible with the joint \texttt{Eureka!} spectrum. While metallicity is not the best description of a rocky planet atmosphere like L 98-59 c, we regardless consider it to facilitate comparison with past works \citepalias[e.g.,][]{Lustigyaeger2023}, and because it is a sensible proxy for atmospheric mean molecular weight. All metallicity-based atmospheres are at chemical equilibrium, which we compute using the open-source Cantera software package \citep{Goodwin2022} with the thermodynamic data used by \citet{Wogan2023} for 91 gas-phase species composed of H, N, O, C, S, and He. Calculations assume a solar C/O ratio, and adopt the \citet{Lodders2009} solar composition. For ``opaque pressure levels'' larger than $10^{-3}$ bar,  the data rules out metallicities $\lesssim 300\times$ solar to 3-$\sigma$ confidence, corresponding to a mean molecular weight of $\lesssim 10$ g mol$^{-1}$ (Figure \ref{fig:sweep}a). For reference, our $300\times$ solar metallicity model has approximately 36\% H$_2$, 31\% H$_2$O, 14\% CH$_4$ and 0.2\% CO$_2$ at the 1 mbar pressure level.

Our new constraint on the metallicity of L 98-59 c shown in Figure \ref{fig:sweep}a is an improvement compared to past estimates that used HST data. We repeated the Figure \ref{fig:sweep}a analysis, but instead using the 6 pixel HST spectrum published in Figure 5 of \citet{barclay_23_arxiv}, which they produced with a single transit observation. This analysis finds that the HST spectrum rules out metallicities $\lesssim 80\times$ solar (MMW $< 4$ g mol$^{-1}$) to 3-$\sigma$ for a 1 bar apparent surface pressure. Our visit 1 and visit 2 joint fit JWST spectrum is more informative, excluding metallicities $\lesssim 300\times$ solar (MMW $< 10$ g mol$^{-1}$) with the same confidence.

We also compare our JWST data to simple H$_2$-CO$_2$ and H$_2$-H$_2$O mixtures as alternative proxies for mean molecular weight (Figures \ref{fig:sweep}b and \ref{fig:sweep}c). For H$_2$-CO$_2$ compositions, the featureless spectrum rules out CO$_2$ concentrations $\lesssim 10\%$ (i.e., MMW $\lesssim 8$ g mol$^{-1}$ ) to 3-$\sigma$ even for low opaque pressure levels (e.g., $10^{-4}$ bar). For H$_2$-H$_2$O mixtures, the data can only exclude H$_2$O mixing ratios $\lesssim 0.5$ (i.e., MMW $< 10$ g mol$^{-1}$) for opaque pressure levels $\gtrsim 3 \times 10^{-2}$ bars. Low water vapor concentrations (e.g., $< 10\%$) and mean molecular weights (e.g., $\sim 3$ g mol$^{-1}$) are permitted by the data for H$_2$-H$_2$O gas combinations so long as there is a high cloud deck or the atmosphere is thin (e.g., $\sim 10^{-3}$ bars; Figure \ref{fig:sweep}c).

The results in Figure \ref{fig:sweep} are qualitatively unchanged for calculations that instead do not permit an offset between the NRS1 and NRS2 detectors. Furthermore, our conclusions are insensitive to the choice of data reduction pipeline. Performing the same analysis with either the \texttt{Eureka!} joint fit or \texttt{ExoTiC-JEDI} weighted average spectrum (Figure \ref{fig:spectra_combined}) produces similar results.

\begin{figure*}
    \centering
    \includegraphics[width=0.7\textwidth]{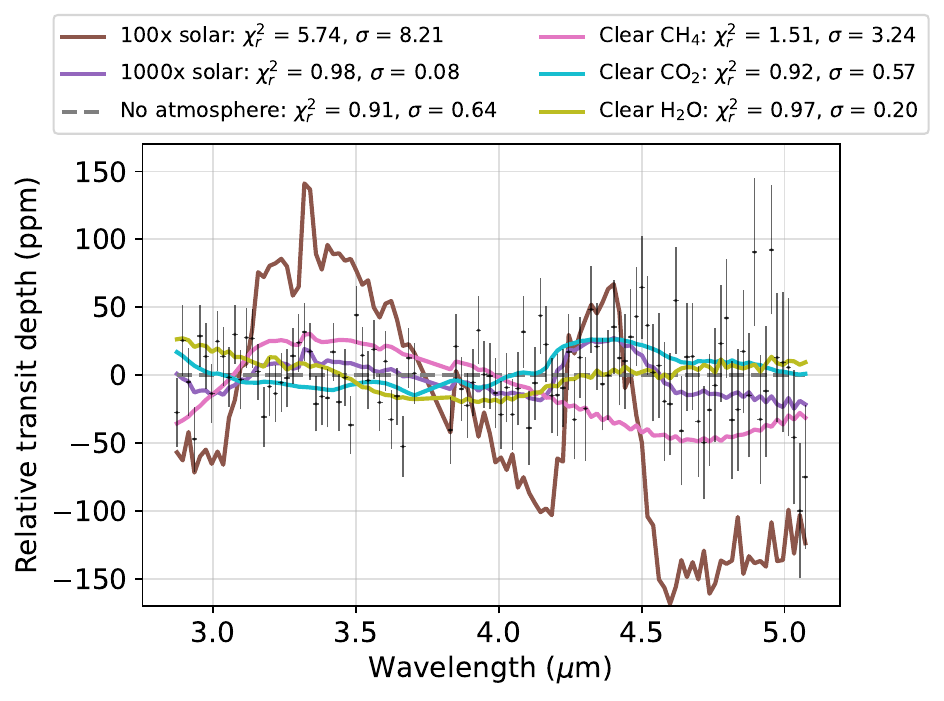}
    \caption{The combined \texttt{Eureka!} spectrum (black dots) compared with models. Zero transit depth is defined by the median of the \texttt{Eureka!} data. The $\chi^2_r$ and $\sigma$ values in the legend are from fits that permit an offset between the NRS1 and NRS2 detectors. The data strongly rule out ($8.2$-$\sigma$) a hydrogen dominated atmosphere with $100 \times$ solar composition, and also disfavors ($2.8$-$\sigma$) a clear CH$_4$ atmosphere. The data are consistent with high mean molecular weight atmospheres such as those with a $1000 \times$ solar, CO$_2$-rich or H$_2$O-rich composition.}
    \label{fig:possible_atmospheres}
\end{figure*}

Figure \ref{fig:possible_atmospheres} further illustrates the atmospheres that are compatible with the transmission spectrum. The data disfavors a clear CH$_4$ atmosphere ($3.2$-$\sigma$), since CH$_4$ has a relatively low mmw and a large feature in the NRS1 bandpass that our precision is sufficient to exclude. Clear-sky CO$_2$ or H$_2$O compositions are permitted, since these molecules have a high mmw and more continuum-like absorption, respectively, in the wavelength regime we are probing. The observations are also possibly explained by a bare rock with no atmosphere (consistent at the $<1$-$\sigma$ level).

\subsection{Interior and Bulk Composition Modelling}\label{subsec:bulkcomposition}
To complement our observational constraints on the atmosphere of L 98-59 c, we also perform interior modelling to infer its volatile content. To do this, we use the open-source code \texttt{smint} \citep{Piaulet2021}\footnote{\url{https://github.com/cpiaulet/smint}}. \texttt{smint} takes planet mass and radius and incident flux as inputs, and uses an MCMC to traverse a pre-computed grid of interior models, returning a posterior distribution for planetary volatile mass for one of two interior scenarios. The first scenario considers a refractory interior (iron core + silicate mantle) of variable core mass fraction with a pure water envelope/atmosphere on top \citep{Aguichine2021}, while the second considers an Earth-like interior with an H$_2$-He envelope of solar metallicity \citep{LopezFortney2014} on top. For the steam-world scenario, the MCMC uses gaussian priors on mass, radius, and equilibrium temperature, and returns the bulk water mass fraction. For this scenario, we predict that planets L 98-59 c and L 98-59 d have bulk water mass fractions of $6.1^{+5.1}_{-3.6}\%$ and $15.6^{+9.1}_{-7.1}\%$, respectively. For the gas dwarf scenario, the MCMC uses Gaussian priors on mass, radius, and incident flux, and a flat prior on age, and returns the bulk H/He mass fraction. For this scenario, we find bulk H/He contents of $0.030^{+0.044}_{-0.016}$\textpertenthousand and $0.10^{+0.13}_{-0.08}$\textpertenthousand, for L 98-59 c and L 98-59 d, respectively.

\begin{figure*}
    \centering
    \includegraphics[width=0.99\textwidth]{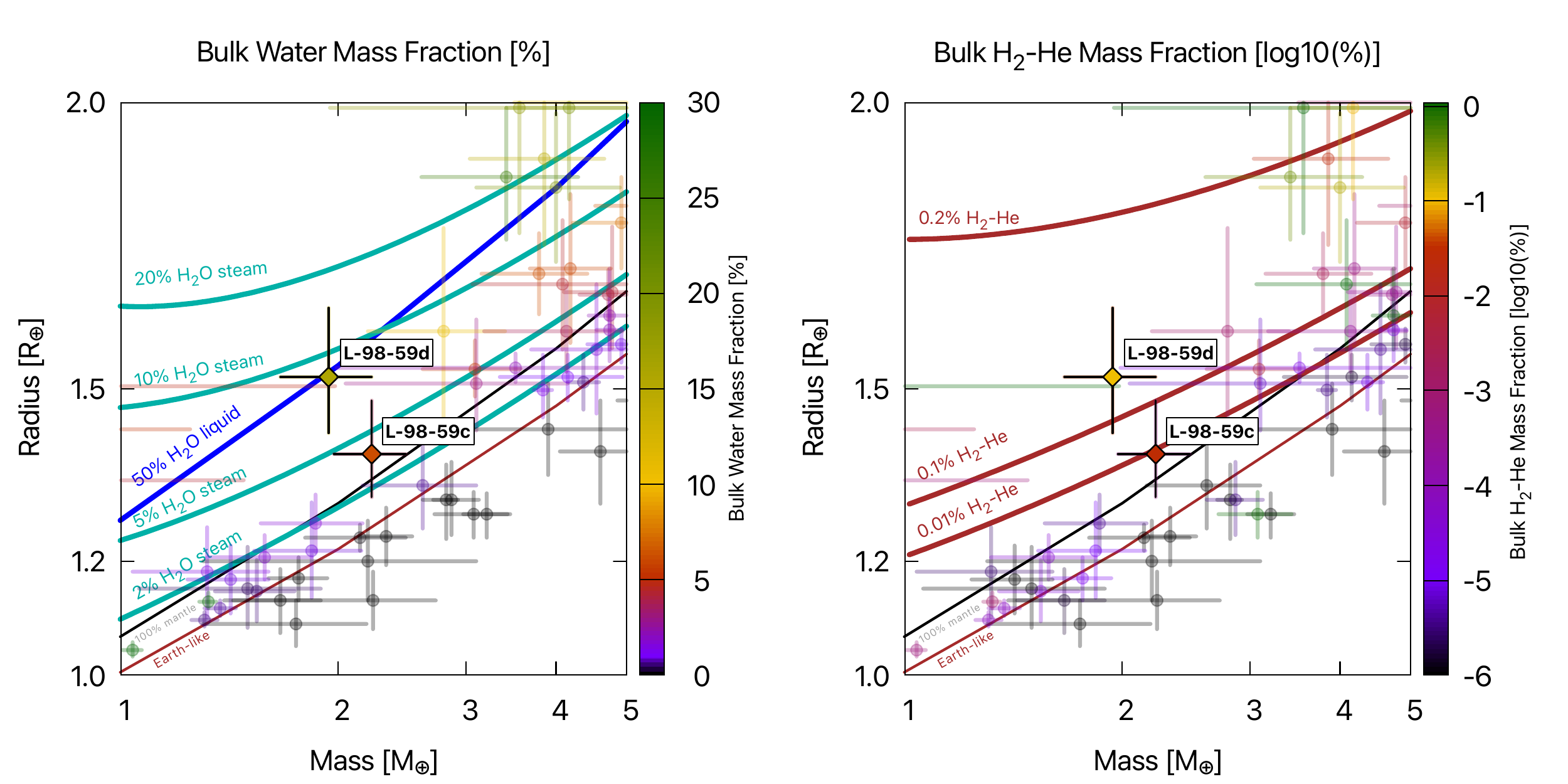}
    \caption{A mass-radius diagram of small exoplanets, colored by bulk water mass fraction (left) and bulk H$_2$/He mass fraction (right). We highlight L 98-59 c, the subject of this study, as well as its outer companion L 98-59 d, which is included in the COMPASS statistical sample, as diamond points. Other planets with mass and radius precision better than 20\% and 5\%, respectively, are shown as circles. Our composition models are drawn from \citet{Zeng2016} for the Earth-like, pure mantle and 50\% liquid water compositions; from \citet{Aguichine2021} for the 2\% to 20\% steam compositions; and from \citet{LopezFortney2014} for the 0.01 to 0.2\% H$_2$-He compositions (assuming a 5 Gyr age and T$_\textrm{eq}=$500K). For each planet, we determine the bulk volatile content using the method described in the text.}
    \label{fig:massradius}
\end{figure*}

Our interior modelling results can be seen in Figure \ref{fig:massradius}. L 98-59 c has a density consistent with either possessing a few percent by mass steam atmosphere, or a $\sim$0.03\% by mass solar metallicity H$_2$-He envelope. 
We consider these end-member cases because there are (as yet) no open source models with coupled atmosphere-interior structure with varying atmospheric metallicity.
The former case is consistent with our observations (Figure \ref{fig:possible_atmospheres}; a pure H$_2$O atmosphere is not ruled out). 
It would also fit within the framework advanced by \citet{luquepalle2022}, namely that the super-Earth population shows dispersion from the Earth-like isocomposition curve in mass-radius space because of inflation by $\sim$ few percent by mass steam envelopes. 
The latter case, of a 0.03\% mass H$_2$-He envelope, is not consistent with our transmission spectrum as presented, i.e. a solar metallicity envelope. 
However, compositions intermediate between the allowed pure H$_2$-He and pure H$_2$O atmosphere, i.e. an envelope with higher metallicity, would also match the bulk properties of L 98-59 c \citep[e.g.,][]{Piaulet_2023}. A subset of these cases are allowed by our transmission spectrum (Figure \ref{fig:sweep}).
Such a higher metallicity scenario could occur due to outgassing of heavier molecular species (see Section \ref{subsec:thickatmosphere}), for example. 
In any case, although this scenario (small H$_2$-He envelope with enhanced metallicity) is notionally consistent with interior models, it is unlikely that this planet (P=3.7d) retains any H$_2$-He envelope due to efficient escape of hydrogen by photoevaporation or boil-off \citep{Rogers2015, Rogers2021, Rogers2023}. 
Based on interior models, L 98-59 c is therefore most likely to be a rocky planet, with a thick atmosphere dominated by high mean molecular weight gases (e.g. H$_2$O and/or CO$_2$), or no atmosphere at all -- conclusions that are all consistent with our transmission spectrum. 

\section{Discussion: Possibilities for L 98-59 \MakeLowercase{c}'\MakeLowercase{s} Atmosphere}\label{sec:discussion}

The 3-5~$\mu$m transmission spectrum of L 98-59 c does not have any discernible features at the level of precision of our observations, and its composition from interior structure modelling does not offer unambiguous conclusions about its atmosphere. We discuss the potential interpretations of these results in the following subsections. 

\subsection{A Thick Atmosphere, Potentially with Clouds/Hazes}\label{subsec:thickatmosphere}

One viable interpretation of L 98-59 c's transmission spectrum is that the atmosphere has a high mean molecular weight and little H/He content -- the latter already suggested by its mass and radius \citep[e.g.,][]{LopezFortney2014, Rogers2015, Zeng2016, Rogers2021}, as well as existing HST observations \citep{Zhou2022, barclay_23_arxiv, Zhou2023}. We rule out a pure CH$_4$ atmosphere at $\sim$2.8$\sigma$. Other potential background gases like H$_2$O and CO$_2$ require mixing ratios $\gtrapprox$40\% and $\gtrapprox$30\% within an H$_2$ atmosphere, respectively, to be compatible with the data, although a very low-pressure $\tau=1$ surface due to clouds or hazes (e.g. 10$^{-4}$ bar) would allow for lower mixing ratios of high-mmw species (see Figure \ref{fig:sweep}). 

A variety of high mean molecular weight atmospheres are thought to be possible for a planet of this size. Spectrally-inactive gases like N$_2$ could play significant roles \citep{Hammond2017,Wakeford2019}, as is the case on Earth, increasing the atmospheric mean molecular weight and depressing feature height. CO$_2$-dominated atmospheres like that of Venus could manifest due to significant outgassing from an oxidized mantle \citep[e.g.,][]{Lebrun2013}. O$_2$-dominated atmospheres could be generated as a result of photolysis of water and subsequent H escape \citep[e.g.,][]{LugerBarnes2015}, a process that could add 15 bars or more of O$_2$ to L 98-59 c's atmosphere \citep{Fromont24}. 
In general, small planets around M stars may not retain significant amounts of atmospheric water \citep[e.g.,][]{Lobo2023}. In the case of L 98-59 c, escape processes are likely to have stripped most of its water within 1 Gyr unless it started with a water inventory of order $\sim$100 Earth oceans \citep{Fromont24}. 
However, our observations do not rule out a H$_2$O-dominated atmosphere (Figure \ref{fig:sweep}c). 

\citet{Seligman2024} recently suggested that tidal forces could melt much of the planetary interior, driving significant volcanism and producing a thick SO$_2$-rich atmosphere for L 98-59 c. We have simulated the transmission spectra of SO$_2$-CO$_2$ mixtures in a 100 bar atmosphere and found that any combination of the gases is permitted by the joint \texttt{Eureka!} spectrum to within $1\sigma$. This aligns with the predictions of \citet{Seligman2024} who find that 3 - 5 transits would be required to detect such compositions (whereas we have only two). Note, however, that a large concentration of volcanic SO$_2$ is perhaps unlikely in a thick atmosphere because sulfur is highly soluble in magma at high pressures. The low solubility of CO$_2$ means its degassing is likely to dominate over SO$_2$, unless the surface atmospheric pressure is very small \citep[e.g., $10^{-3}$ bar;][]{Gaillard2014}.

%

\subsection{No Atmosphere}

The other interpretation consistent with our data is that this planet has no (or a very tenuous) atmosphere. Modelling undertaken to understand small planets around M stars suggests that strong XUV fluxes from the host star could completely strip planetary atmospheres \citep{ZahnleCatling2017}. Recent studies with JWST's Mid-Infrared Instrument (MIRI) have suggested that TRAPPIST-1 b and c both lack significant atmospheres \citep{Greene2023,Zieba2023}. These studies serve as preliminary evidence that moderately-irradiated ($T_\mathrm{eq} \sim 400$ K), low-mass planets around M stars, like L 98-59 c (which receives roughly three times the irradiation of TRAPPIST-1 b), may all have difficulty holding onto atmospheres. While the density of L 98-59 c is suggestive of an atmosphere (Section \ref{subsec:bulkcomposition}), the TRAPPIST-1 planets are similarly under-dense relative to an Earth-like composition \citep{Agol2021}. This could be due to pure silicate compositions \citep[i.e. core-free;][]{ElkinsTanton2008}, a scenario which neither our spectrum nor our interior modelling can rule out for L 98-59 c. Thus, the no atmosphere scenario remains a plausible explanation for our observations. 

\subsection{Thermal Emission Observations Should Be Able To Detect a Thick Atmosphere}

At the time of writing of this manuscript, JWST has observed a secondary eclipse of L 98-59 c using the F1500W MIRI filter, but the data has not yet been published (GO 3730, PI: Diamond-Lowe). The resulting constraints on the planet's dayside thermal emission will be sensitive to the presence or absence of an atmosphere. If L 98-59 c does not have an atmosphere then its dayside emission should be relatively large because all absorbed starlight should be instantly re-radiated \citep{Koll2022}. Conversely, a thick atmosphere would exhibit comparatively low day side emission because winds can redistribute absorbed stellar energy to the night side. 

\begin{figure*}
    \centering
    \includegraphics[width=0.8\textwidth]{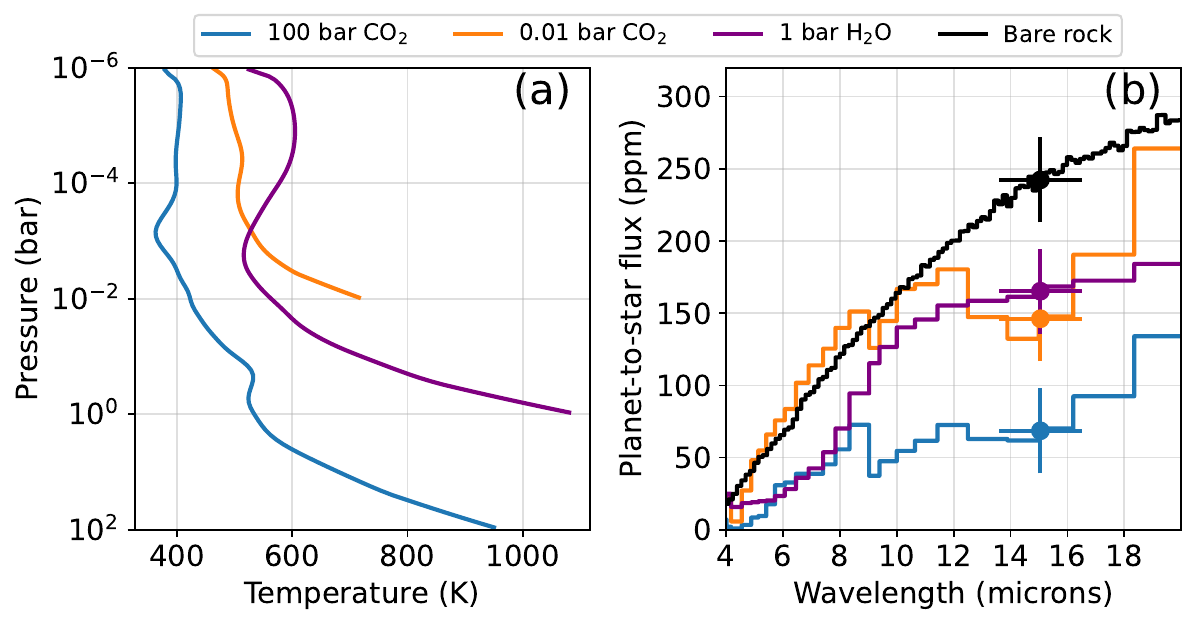}
    \caption{Predicted day side thermal emission from L 98-59 c. Panel (a) shows computed day side pressure-temperature profiles for three possible atmospheric compositions: 100 bars of CO$_2$ (blue), 0.01 bars if CO$_2$ (orange), and 1 bar of steam (purple). Panel (b) gives the day side planet-to-star flux ratio of each atmosphere compared to a bare rock with a 0.1 bond albedo (black). Each scenario in (b) includes synthetic JWST/MIRI eclipse measurements at 15 $\mu$m with expected 30 ppm uncertainty. If L 98-59 c has a CO$_2$ atmosphere exceeding $\sim 0.01$ bars or a steam atmosphere $\gtrsim 1$ bar, then a 15 $\mu$m MIRI thermal emission observation will rule out a bare rock scenario to $\gtrsim 3\sigma$ confidence.}
    \label{fig:thermalemission}
\end{figure*}

To illustrate the potential effectiveness of this new secondary eclipse observation, we compute synthetic F1500W MIRI data for a variety of possible atmospheres and estimate whether they can be confidently distinguished from a bare rock scenario. To simulate dayside climate and emission we use the open-source 1D climate model in \citet{Wogan2023} and account for day-night heat redistribution following the parameterization in \citet{Koll2022}. Figure \ref{fig:thermalemission} shows the computed day side pressure-temperature profiles and thermal emission spectra for four atmospheric cases compatible with our transmission spectrum: 100 bars of CO$_2$, 0.01 bars if CO$_2$, 1 bar of steam, and a bare rock assuming a bond albedo similar to Mercury’s ($\alpha_B = 0.1$). Each scenario in Figure \ref{fig:thermalemission}b includes a synthetic JWST/MIRI eclipse measurement at 15 $\mu$m with 30 ppm error, which is the precision predicted by the JWST exposure time calculator \citep{Pontoppidan2016} for a single secondary eclipse observation. If L 98-59 c has a Venus-like 100 bar CO$_2$ atmosphere, then the 15 $\mu$m MIRI observation will rule out a bare rock with 5.9$\sigma$ confidence. For a 0.01 bar CO$_2$ atmosphere or 1 bar steam atmosphere, a bare rock is instead disfavored by 3.3$\sigma$ or 2.6$\sigma$, respectively.

We also considered possible atmospheric compositions for L 98-59 c beyond those shown in Figure \ref{fig:thermalemission}. For example, an atmosphere dominated by CO, O$_2$, or N$_2$ would be detectable for pure compositions exceeding 10s to 100s of bars, or for smaller pressures if the atmosphere also contains $\gtrsim$0.01\,bar of CO$_2$. We do not consider H$_2$- or CH$_4$-rich compositions because they are disfavored by our transmission spectrum (e.g., Figure \ref{fig:possible_atmospheres}). We may soon have further evidence to narrow down the compositional parameter space in which this fairly cool ($T_{eq} \approx 550$~K) M dwarf super-Earth resides.

\section{Summary}\label{sec:summary}

A summary of our paper's key points are as follows. 

\begin{itemize}
    \item We have presented the 3-5 micron JWST NIRSpec G395H transmission spectrum of L 98-59 c, a warm ($\sim$550K) super-Earth in a nearby system. 
    \item Two independent reductions agree that the planet's transmission spectrum does not show features indicative of molecular absorption at the precision of our observations (median uncertainty 28.6ppm).
    \item Using this apparently featureless transmission spectrum, we rule out atmospheric metallicities below $\sim$300$\times$ solar (assuming opaque pressure levels $\gtrsim$1mbar), or correspondingly atmospheric mean molecular weights below $\sim$10~g/mol. We also rule out clear-sky pure methane atmospheres. 
    \item The remaining atmospheric parameter space allowed 
    for this planet includes high mean molecular weight atmospheres (e.g., H$_2$O- or CO$_2$-rich), lower mean molecular weight compositions with very low opaque pressure levels (e.g., due to high-altitude clouds or a thin atmosphere), or no atmosphere at all.
\end{itemize}

The constraints laid out here are only the beginning for JWST studies of the L 98-59 system. Other programs will view planet c in emission (GO 3730, PI: Diamond-Lowe) and transmission with a different instrument (NIRISS; GTO 1185, PI: Lafreniere). Planet d will also be studied in transmission with both NIRSpec (GTO 1224, PI: Birkmann) and NIRISS (GTO 1185, PI: Lafreniere). Because of this system's multiplicity and bright host star, it will likely be the subject of additional future studies, and may be a key part of our understanding of the atmospheres of terrestrial planets around M dwarfs. 

\section{Acknowledgments}

We thank the anonymous referee for their comments, which helped improve the quality and clarity of this manuscript. This work is based on observations made with the NASA/ESA/CSA James Webb Space Telescope. Data products for this manuscript can be found in Zenodo at doi:10.5281/zenodo.13363863.
The data were obtained from the Mikulski Archive for Space Telescopes at the Space Telescope Science Institute, which is operated by the Association of Universities for Research in Astronomy, Inc., under NASA contract NAS 5-03127 for JWST. 
These observations are associated with program \#2512. Support for program \#2512 was provided by NASA through a grant from the Space Telescope Science Institute, which is operated by the Association of Universities for Research in Astronomy, Inc., under NASA contract NAS 5-03127. 
All of the JWST data used for this work can be found at \dataset[10.17909/qdea-wg88]{http://dx.doi.org/10.17909/qdea-wg88}.
NS gratefully acknowledges support from the Heising-Simons Foundation through grant 2021-3197.
This work is funded in part by the Alfred P. Sloan Foundation under grant G202114194. Support for this work was provided by NASA through grant 80NSSC19K0290 to JT and NW. 
This work benefited from the 2022 and 2023 Exoplanet Summer Program in the Other Worlds Laboratory (OWL) at the University of California, Santa Cruz, a program funded by the Heising-Simons Foundation. 
This material is based upon work supported by NASA’S Interdisciplinary Consortia for Astrobiology Research (NNH19ZDA001N-ICAR) under award number 19-ICAR19\_2-0041. 
H.R.W. was funded by UK Research and Innovation (UKRI) under the UK government’s Horizon Europe funding guarantee for an ERC Starter Grant [grant number EP/Y006313/1].
This research has made use of the NASA Exoplanet Archive, which is operated by the California Institute of Technology, under contract with the National Aeronautics and Space Administration under the Exoplanet Exploration Program \citep{nea}.

Co-author contributions are as follows: NS led the data analysis with contributions from NLW. NW led the interpretation with contributions from NS and NEB. HRW provided data reduction and analysis with a second pipeline with contributions from LA. AA provided interior models. All authors provided comments and discussion over the course of the analysis and detailed feedback on the manuscript.

\software{\texttt{emcee} \citep{ForemanMackey2013}, \texttt{Eureka!} \citep{Bell2022_eureka},  \texttt{ExoTiC-Jedi} \citep{Alderson2022_jedi_zenodo}, \texttt{ExoTiC-LD} \citep{Grant2022}, \texttt{Matplotlib} \citep{Hunter2007},  \texttt{NumPy} \citep{Harris2020}, \texttt{PandExo} \citep{Batalha_2017}, \citep{Virtanen2020}, \texttt{PICASO} \citep{Batalha2019_picaso}, \texttt{Photochem.clima} \citep{Wogan2023},  \texttt{smint} \citep{Piaulet2021}, \texttt{ultranest}\citep{Buchner2021}}


%




\bibliography{main.bib}
\bibliographystyle{aasjournal}


\begin{appendix}
\setcounter{figure}{0}

\section{Our Conclusions Are Not Significantly Affected by Choice of Fixed or Fit Astrophysical Parameters} \label{sec:appendix-astro-param}

As discussed in Main Text Section \ref{subsec:fitvsfixed}, there is some disagreement between reductions, detectors, and visits in the best fit astrophysical parameters to our white light curves. However, we show here that our choice to fix the inclination in our fitting does not significantly impact the conclusions drawn from our spectrum (i.e., that it is featureless). We attempted fits with all parameters fixed to the values in \citet{Demangeon21}, but as can be seen in Appendix Figure \ref{fig:appendix_lcs}, these parameters do not provide a good fit to our observed light curves. Fixing the inclination only results in consistent a/R$_\star$ values as well as good quality light curve fits (Main Text Table 3; Main Text Figure \ref{fig:combined_wlcs}), so we adopt that model. That choice, however, does not impact the resulting spectrum. In Appendix Figure \ref{fig:appendix_spectrum}, we show the \texttt{Eureka!} spectrum for Visit 2 (most strongly impacted by best-fit white light curve parameter differences) generated by fixing or fitting for inclination. These spectra clearly agree with each other at better than 1$\sigma$. For brevity we show only the \texttt{Eureka!} reductions here but similar effects emerge in the \texttt{ExoTiC-JEDI} reductions. Future work might apply additional techniques to mitigate the impact of correlated noise, but this effort is beyond the scope of our work here, and the agreement between spectra produced with these two techniques demonstrates that our conclusions are robust to the details of the white light curve fitting. 

\begin{figure*}[h]
    \centering
    \includegraphics[width=0.99\textwidth]{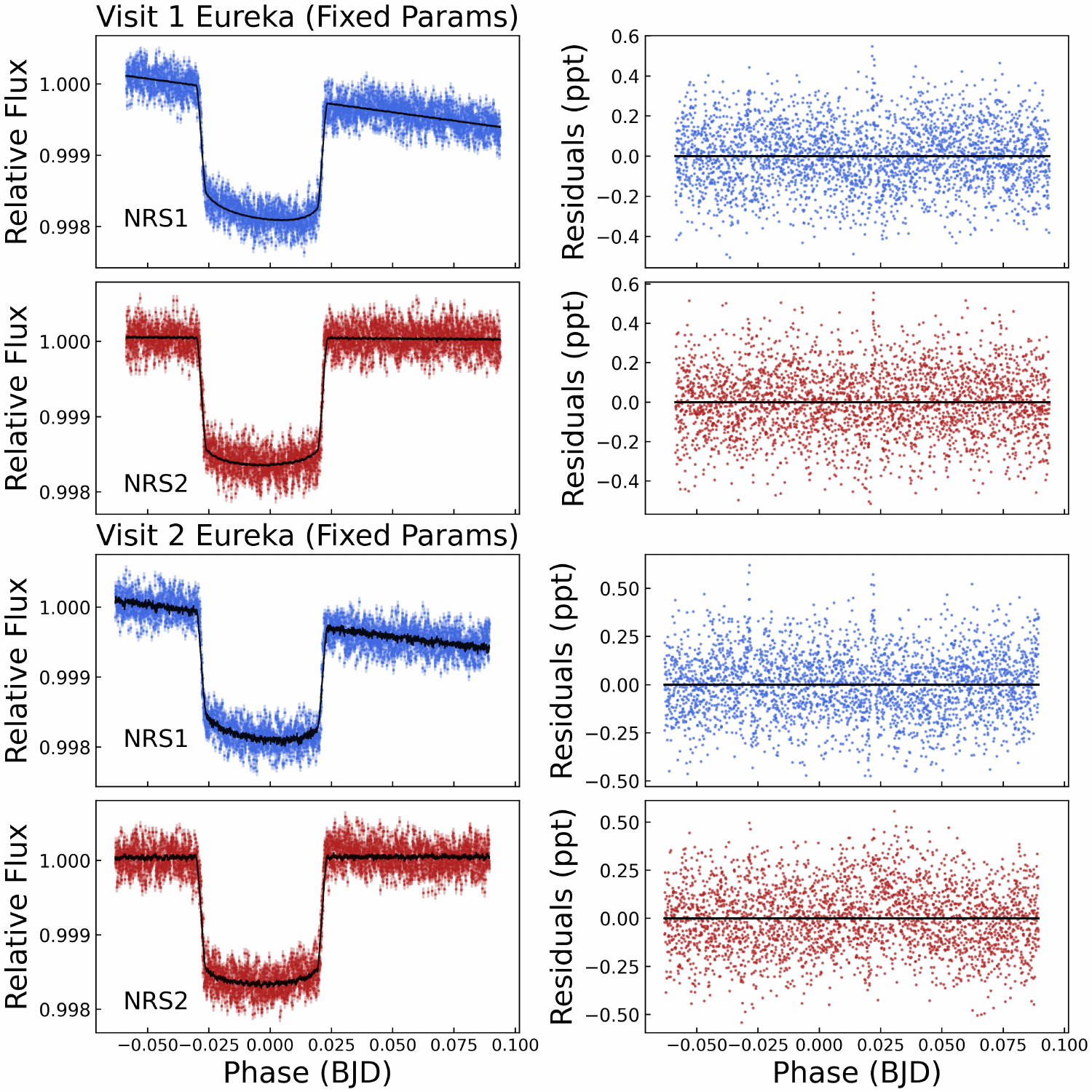}
    \caption{Left: The L 98-59 c white light curves from \texttt{Eureka!} for each detector and visit. The best-fit models with a/R$_*$ and inclination fixed to the values from \citet{Demangeon21}, including systematics, are overplotted on the data. Right: the residuals from the best-fit models. Unlike the fits where only inclination is fixed, the residuals display clear spike-like structure near ingress and egress, suggesting these values are not a good fit to our data.}
    \label{fig:appendix_lcs}
\end{figure*}

\begin{figure*}[h]
    \centering
    \includegraphics[width=0.99\textwidth]{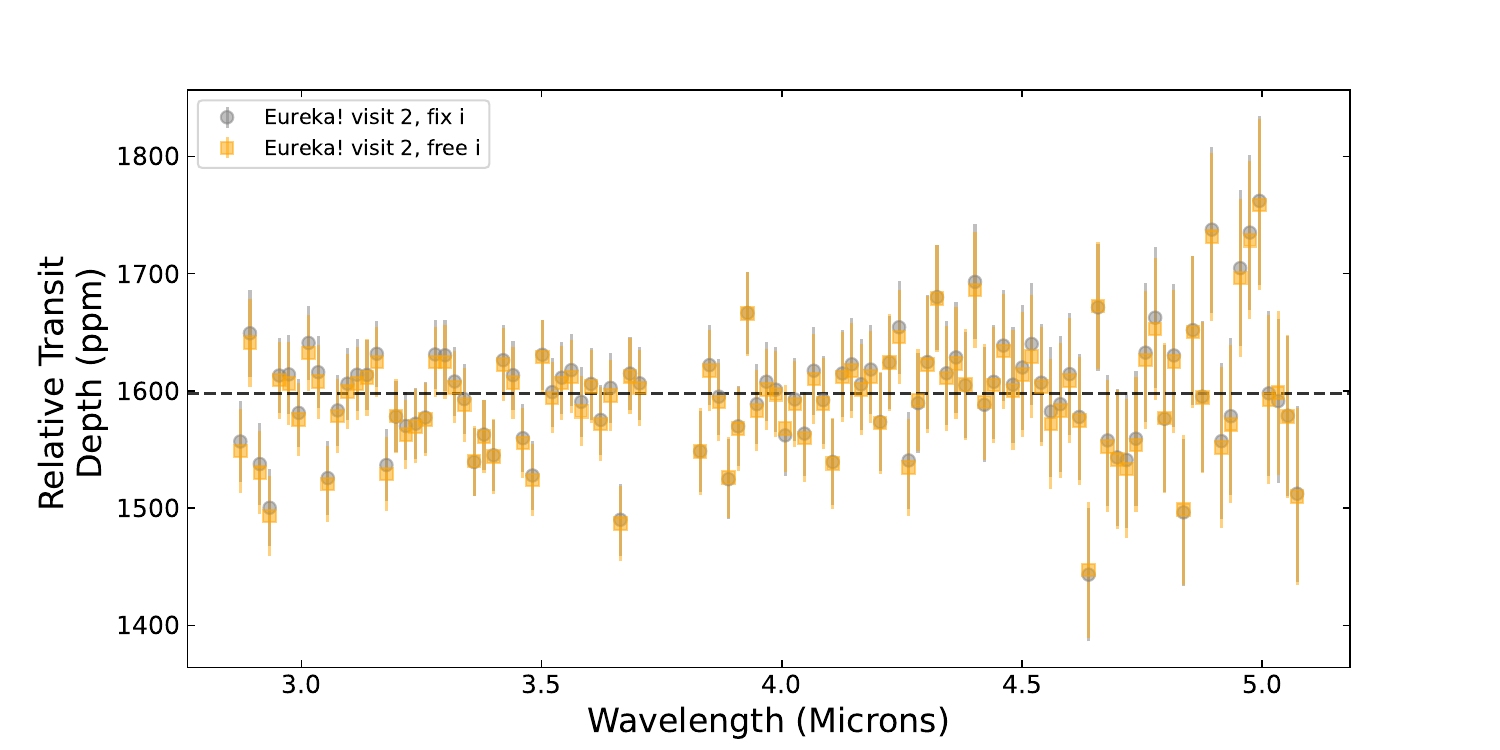}
    \caption{The \texttt{Eureka!} spectrum resulting from the second visit of L 98-59 c depending on whether the inclination is fixed or free.}
    \label{fig:appendix_spectrum}
\end{figure*}

\end{appendix}

\end{document}